\documentclass[twocolumn,trackchanges]{aastex62}
\newcommand{\ddo}{$D_{0.2}$}
\newcommand{\hgamma}{$\mathrm{H}\gamma$}
\newcommand{\hdelta}{$\mathrm{H}\delta$}
\newcommand{\hbeta}{$\mathrm{H}\beta$}
\newcommand{\ffm}{$f_m$}

\received{January 1, 2022}
\revised{January 7, 2022}
\accepted{\today}
\submitjournal{ApJs}

\shorttitle{BHB stars}
\shortauthors{Ju et al.}
\usepackage{amsmath}
\begin{document}

\title{Identification of Blue Horizontal-Branch Stars From LAMOST DR5}

\correspondingauthor{Wenyuan Cui; Chao Liu}
\email{wenyuancui@126.com, cuiwenyuan@hebtu.edu.cn; liuchao@bao.ac.cn}

\author{Jie Ju}
\affil{Department of Physics, Hebei Normal University, Shijiazhuang 050024, China}
\affil{School of Sciences, Hebei University of Science and Technology, Shijiazhuang 050018, China}

\author[0000-0003-1359-9908]{Wenyuan Cui}
\affil{Department of Physics, Hebei Normal University, Shijiazhuang 050024, China}

\author{Zhenyan Huo}
\affiliation{Department of Physics, Hebei Normal University, Shijiazhuang 050024, China}

\author{Chao liu}
\affiliation{Key Laboratory of Space Astronomy and Technology, National Astronomical Observatories, Chinese Academy of Sciences, Beijing 100101, China}

\author{Xiangxiang Xue}
\affiliation{Key Laboratory of Optical Astronomy, National Astronomical Observatories, Chinese Academy of Sciences, Beijing 100101, China}

\author{Jiaming Liu}
\affiliation{Department of Physics, Hebei Normal University, Shijiazhuang 050024, China}

\author{Shuai Feng}
\affiliation{Department of Physics, Hebei Normal University, Shijiazhuang 050024, China}

\author{Mingxu Sun}
\affiliation{Department of Physics, Hebei Normal University, Shijiazhuang 050024, China}

\author{Linlin Li}
\affiliation{Department of Physics, Hebei Normal University, Shijiazhuang 050024, China}
%
%



\begin{abstract}
We construct a new catalog
of the blue horizontal-branch (BHB) stars from the   Large Sky Area Multi-Object Fiber Spectroscopic Telescope (LAMOST) DR5 dataset, which contains 5355+81 BHB stars at high Galactic latitude ($|Glat|>20^{\circ}$). We combine the spectral line indices with a set of Balmer line profile selection criteria to identify the BHB stars. During the selection process, we use the line index of \ion{Ca}{2}\,K to exclude the metal-rich A-type dwarfs. We obtain their atmospheric parameters by cross-matching our BHB stars with the catalog provided by \citet{Xiang2022}. The results show that our sample is consistent with the theoretical $T_{\rm eff}$-log\,$g$ evolutionary tracks of the BHB stars, indicating that our method is robust for identifying BHB stars from the LAMOST spectra.  
Their spatial distribution indicates that most of our BHB stars are located in the inner halo or the disk of the Milky Way. Combined with other BHB samples from the literature, the BHB stars can cover a large Galactic volume, which makes it a better probe for studying the kinematics, dynamics, and structural characteristics of the Milky Way.

\end{abstract}



\section{Introduction} 

The blue horizontal-branch (BHB) stars are old, metal-poor Population II stars with mass less than 1.0\,$M_\odot$, which are often found in the Galactic halo. They are first clearly detected present day and well-defined in Galactic globular clusters \citep{Arp1952}. BHB stars generally rotate slowly, with $vsini<$ 20 km s$^{-1}$, but some stars are found to be rotating as high as 40 km s$^{-1}$ \citep{Cohen1997}.
 BHB stars are located on the blue side of RR Lyrae variables on the Hertzsprung-Russell (H-R hereafter) diagram. It now burns helium in its core and hydrogen in shells outside the core \citep{Ruhland2011}. The majority of the BHB stars are A-type stars \citep[$T_{\rm eff}<$ 12,000 K,][]{Catelan2009}. 

BHB stars are luminous and have a nearly constant absolute magnitude within a restricted color range, allowing them to be identified even at large distances. Historically, BHB stars and RR Lyrae stars as the standard candle for tracing the distance are usually used to constrain the kinematics of the local halo \citep{Pier1984, Layden1996}. Considering that the number of BHB stars is more than ten times that of RR Lyrae stars, it is more likely to use BHB stars to study Galactic halo \citep{Pier1983, Sommer1986, Flynn1988, Sommer1989, Preston1991, Arnold1992, Kinman1994, Beers1996, Beers2007, Christlieb2007, Thom2005, Sommer1997, Preston1991, York2000, Santucci2011, Deason2011, Clewley2006}, and they are
the most favorable for estimating the mass distribution of the halo. Based on the sample of BHB stars from the Sloan Digital Sky Survey 
 \citep[SDSS,][]{York2000}, \citet{Xue2011} accurately determined the Galactic mass, i.e.,  about 100 billion solar masses.

Early catalogs \citep{Pier1983, Beers1988} relied mostly on the data obtained by the narrow-band objective-prism surveys, where two main features of strong  $\mathrm{H}\theta$ and weak or absent \ion{Ca}{2}\,K were used. Two main methods have been used to identify BHB stars from the data gathered by large-scale sky surveys, one of which uses photometric methods. \citet{Beers2007} used the coordinates and photometric data ($-0.2\le (B-V)_0\le0.2$)  to identify 12,056 field horizontal-branch stars from the Two Micron All Sky Survey \citep[2MASS,][]{Skrutskie2006}. \citet{Brown2008} selected 2,414 BHB candidates from 2MASS catalog using 
$ -0.2< (J-H)_0<0.1$, 
$-0.1<(H-K)_0<0.1$ and $12.5 < J_0 < 15.5$ with 67\% purity. \citet{Montenegro2019} discovered 12,554 BHB stars using the globular cluster M22 as a reference standard and constructing color–magnitude, and color–color diagrams with precise cuts in the ZYJHK from the Vía Láctea (VVV) ESO Public Survey data. \citet{Culpan2021} selected 57,377 BHB stars using the color and absolute magnitude ($G, G_{BP}-G_{RP}$) from Gaia early data release 3 \citep[EDR3,][]{Gaia2020}, and the purity is about 70{\%}. 
But the effect of extinction is inevitable, which can usually lead to serious
contamination. Another approach is to base not only on photometry but also on spectral features. \citet{Xue2011} got 4,985 BHB stars from SDSS DR8 through color cuts and selection criteria for Balmer line profiles. In addition to the above two methods, \citet{Vickers2021} generated a catalog of 13,693 BHB stars from the LAMOST DR5 through a machine-learning algorithm with the training data from \citet{Brown2008} and \citet{Xue2008}, respectively.

The major difficulty in selecting BHB stars is distinguishing them from  A-type dwarfs and blue stragglers (BS). A-type dwarfs have high surface gravity, while BHB stars have lower surface gravity and show solid and sharp in the Stark pressure-broadened Balmer line profiles \citep{Pier1983, Kinman1994, Wilhelm1999a, Clewley2002, Clewley2006, Brown2008}. BS has similar effective temperatures but is dimmer and also has higher gravity than BHB stars. Photometric methods cannot accurately obtain these features, and spectroscopic methods are necessary. Additionally, BHB stars are metal-poor stars. To avoid the contamination of metal-rich stars, it is able to acquire BHB stars by cutting metallicity.

This work aims to use LAMOST DR5 to construct a new  BHB star catalog with a low contamination rate and extensive coverage of the sky area. 
This paper is structured as follows: In Section 2, we present the primary data of LAMOST DR5 and describe the details of the approach to identifying BHB stars. In Section 3, we investigate the reliability of our sample and discuss our results. The summary and conclusions are in Section 4. 

\section{DATA}

To identify BHB stars, we adopt the method used by \citet{Sirko2004} and \citet{Xue2008} with partial modification. Here, we use the spectral indices cuts instead of their color cuts.
The advantage of the spectral indices over colors is that the effect of Galactic extinction can be avoided.

\subsection{The LAMOST data}
LAMOST is a 4 m reflective Schmidt telescope with a wavelength coverage of 370 - 900\,nm and $R\sim1800$ \citep{Cui2012, Zhao2012, Luo2012}. Due to its unique design, LAMOST can take about 4,000 spectra with a limiting magnitude of $r=18$  \citep{Deng2012}. More than 10 million low-resolution spectra have been obtained. With such a large amount of spectra, many interesting but relatively rare stellar objects have been identified, such as carbon stars with strong CH, CN, and C2 absorption bands \citep{Ji2016, Li2018}, Mira variables with high excitation emission lines which vary with the pulsation cycle \citep{Yao2016}, and OB-type stars with strong neutral or ionized He lines and slightly weak Balmer lines \citep{Liu2019a}, etc. The large stellar spectra sample is also suitable for identifying BHB stars.

%

\subsection{Line indices}

As shown by \citet{Liu2015}, some spectral line indices are sensitive to the atmospheric parameters of stars, e.g., the effective temperature, surface gravity, etc. This means that we can use spectral line features to identify the stars with different spectral types and luminosity classes. Using the criteria of the spectral line indices, carbon stars \citep{Ji2016} and OB-type stars \citep{Liu2019a} have been successfully identified from the LAMSOT DR2 and DR5 data, respectively.

\begin{deluxetable}{lll}
\tablecaption{The definition of the line indices. \label{Table1}}
\tablehead{
\colhead{Name} & \colhead{Index Bandpass (\AA)} &
\colhead{Pseudo-continua (\AA)}}
\startdata
\ion{Ca}{2}\,K \ &3927.70–3939.70&3903.00–3923.00 
 4000.00–4020.00 \\
$\rm G$4300 &4282.62-4317.62&4267.62-4283.87  4320.12-4336.37\\
\hgamma\ &4319.75–4363.50& 4283.50–4319.75  4367.25–4419.75 \\
\enddata
\end{deluxetable}

The line indices in terms of equivalent width ($\mathrm{EW}$) are defined by the following equation:
\begin{equation}\label{eq:ew}
EW=\int(1-\frac{F_\lambda}{F_C})d\lambda.
\end{equation}
where $F_\lambda$ and $F_C$ are the fluxes of the spectral line and the pseudo-continuum, respectively \citep{Worthey1994, Liu2015}. $F_C$ is estimated via linear interpolation of the fluxes located in the ``shoulder'' region on either side of the line bandpass. The unit of line indices under this definition is in \AA. 

According to equation (\ref{eq:ew}), we calculate the $\mathrm{EW}$ of spectral lines listed in Table 1 \citep{Liu2015} for the LAMOST DR5 data.

\section{Identification of BHB stars}


We first exclude the spectra with a signal-to-noise ratio less than ten at the $g$ band (S/N)$_g$ from LAMOST DR5, then 5,807,771 stellar spectra left. Typically the BHB stars have low metallicities and are usually found in the Galactic halo region. To avoid contamination by large numbers of young Galactic disk stars, we select the stars with absolute values of the Galactic latitude (hereafter $Glat$) larger than $20 ^{\circ}$ and obtain 3,490,485 stellar spectra. It should be noted that there are some stars which are been observed multiple times. Of course, with the $Glat$ cuts, we will miss some BHB stars in low Galactic latitude.

\subsection{$\mathrm{EW}$ cuts of H$\gamma$ and G4300}

\begin{figure}
	\epsscale{1.0}
	\plotone{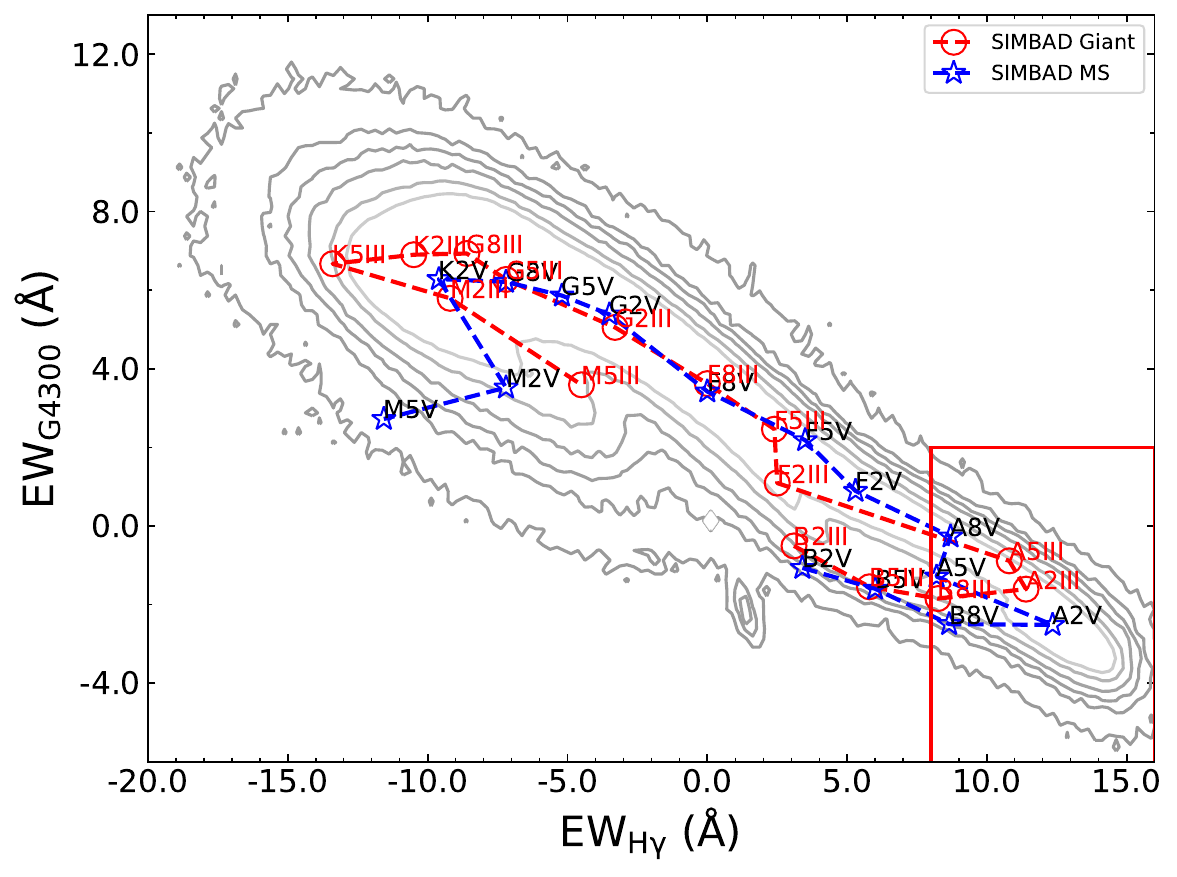}
	\caption{The contours of the LAMOST DR5  spectra with  (S/N)$_g>10$ and $|Glat|>20^{\circ}$} on the $\mathrm{EW_{G4300}}$ vs. $\mathrm{EW_{H\gamma}}$ plane. The blue dashed line with the asterisks and the red dashed line with the unfilled circles represent the loci of the main sequence and giant stars provided by \citet{Liu2015}, respectively. The red solid lines are the line indices' cuts of \hgamma\ at 4341\,\AA\ and $\mathrm{G4300}$ around 4300\,\AA\ for the selection of BHB candidates (mainly A-type stars). \label{fig:hgamma-g4300}
\end{figure}

Here, we mainly focus on the BHB stars with the spectral type of A. Their spectra are characterized by strong hydrogen lines and weak or no molecular bands, which could help to remove the contaminants hotter or cooler than the A-type stars. We plot the above 3,490,485 stars with $|Glat|>20^{\circ}$ on $\mathrm{EW_{G4300}}$ vs. $\mathrm{EW_{H\gamma}}$ plane (see Figure \ref{fig:hgamma-g4300}). Based on the loci of the A-type main sequence (MS, hereafter) and giant stars provided by \citet{Liu2015}, we find that A-type stars are mostly located in the area with smaller values of G4300 but larger values of \hgamma. Therefore, we empirically choose the region enclosed by the red solid line to cover almost all of the A-type stars, especially the stars brighter than the A-type giants. The line indices cuts adopted are as follows: 
\begin{equation}\label{eq:line-ind}
\mathrm{EW_{G4300}}<2.0,
\mathrm{EW_{H\gamma}}>8.0.
\end{equation}
With this cut, we obtain 39,997 spectra as the BHB candidates, in which the contaminants are mainly BS, A-type MS stars, and a few late B- and early F-type stars (see Figure \ref{fig:hgamma-g4300}).

\subsection{Metallicity cuts}
For the 39,997 BHB candidates, we have removed most of the young and metal-rich disk stars with cuts of 
$|Glat|>20^{\circ}$. The \ion{Ca}{2}\,K absorption lines, which are also prominent features in the LAMOST spectra, are often used as an ideal indicator of the stellar metallicity for the low resolution and low S/N spectra \citep{Pier1983, Beers1992, Clewley2002}. 

\begin{figure}
	\epsscale{1.0}
	\plotone{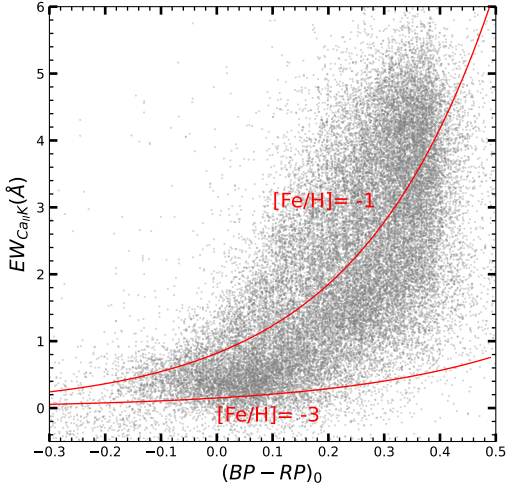}
	\caption{The distributions of 39,445 BHB star candidates with  (S/N)$_g>10$ and $|Glat|>20^{\circ}$ on the $\mathrm{EW_{Ca\,II K}}$ vs. $(BP-RP)_0$ plane. The red solid lines represent $\mathrm{EW_{Ca\,II K}}$ as a function of $(BP-RP)_0$ for stars with [Fe/H]$= -1$ and $-3$, respectively, which are adopted from  \citet{Wilhelm1999a}. \label{fig:Cak-feh-1}}
\end{figure}

To further remove the contaminants with metallicities larger than those of BHB stars, we calculate the $\mathrm{EW}$ of the \ion{Ca}{2}\,K line according to equation~(\ref{eq:ew}). We cross-match our 39,997 BHB candidates with the Gaia EDR3 catalog \citep{Gaia2020} and obtain their $BP$ and $RP$ magnitudes, except for 552 candidates. There are three reasons why these stars do not match with Gaia. One is that some stars are too bright so that the fiber of LAMOST will not target the center of stars. The second one is that there are two stars so close together that it is indistinguishable from each other. The last one is that maybe the star is too faint or located in a crowded area, some stars are not found in Gaia. Figure~\ref{fig:Cak-feh-1} shows the $\mathrm{EW_{Ca\,II\,K}}$ versus $(BP-RP)_0$ for our 39,445 BHB candidates, where $(BP-RP)_0$ adopt the extinction law of \citet{Wang2019}. The reddening values are obtained from \citet{Schlegel1998}. It is generally accepted that the metallicity of most BHB stars is less than -1 \citep{Kinman2000}. In Figure~\ref{fig:Cak-feh-1}, We also plot the theoretical    $\mathrm{EW_{Ca\,II\,K}}$ and $(BP-RP)_0$ relations (red solid lines)  calculated from the synthetic spectra with a resolution close to that of LAMOST spectra and the atmospheric parameters log $g = 3.5$, [Fe/H]$ = -1$ and $-3$\citep{Wilhelm1999a}. Based on the red solid line with [Fe/H]$= -1$, we select the sample with [Fe/H]$\le -1$ and obtain 23,528+552 BHB candidates. 

\subsection{Balmer line profile cuts}

\begin{figure}
	\epsscale{1.2}
	\plotone{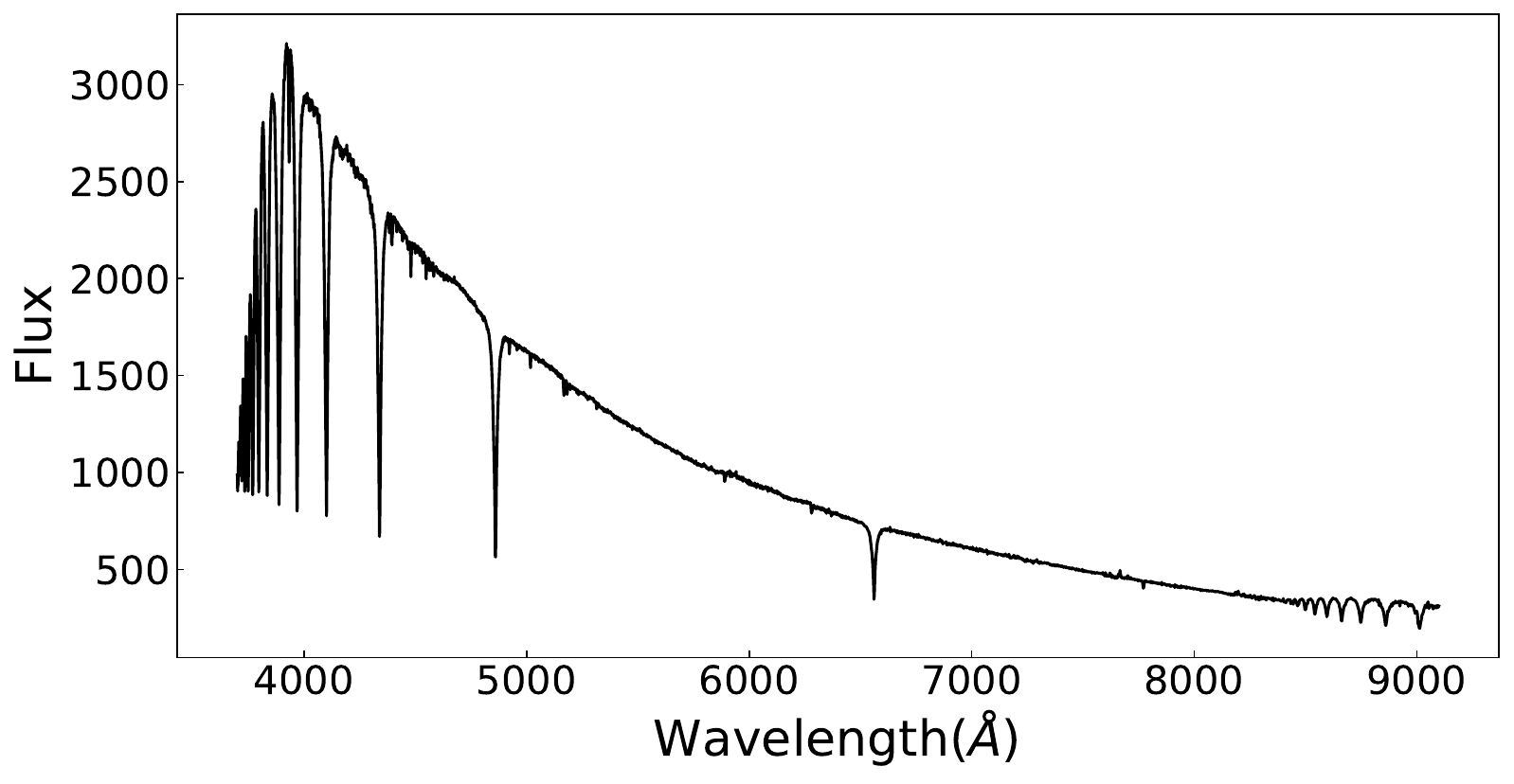}
	\plotone{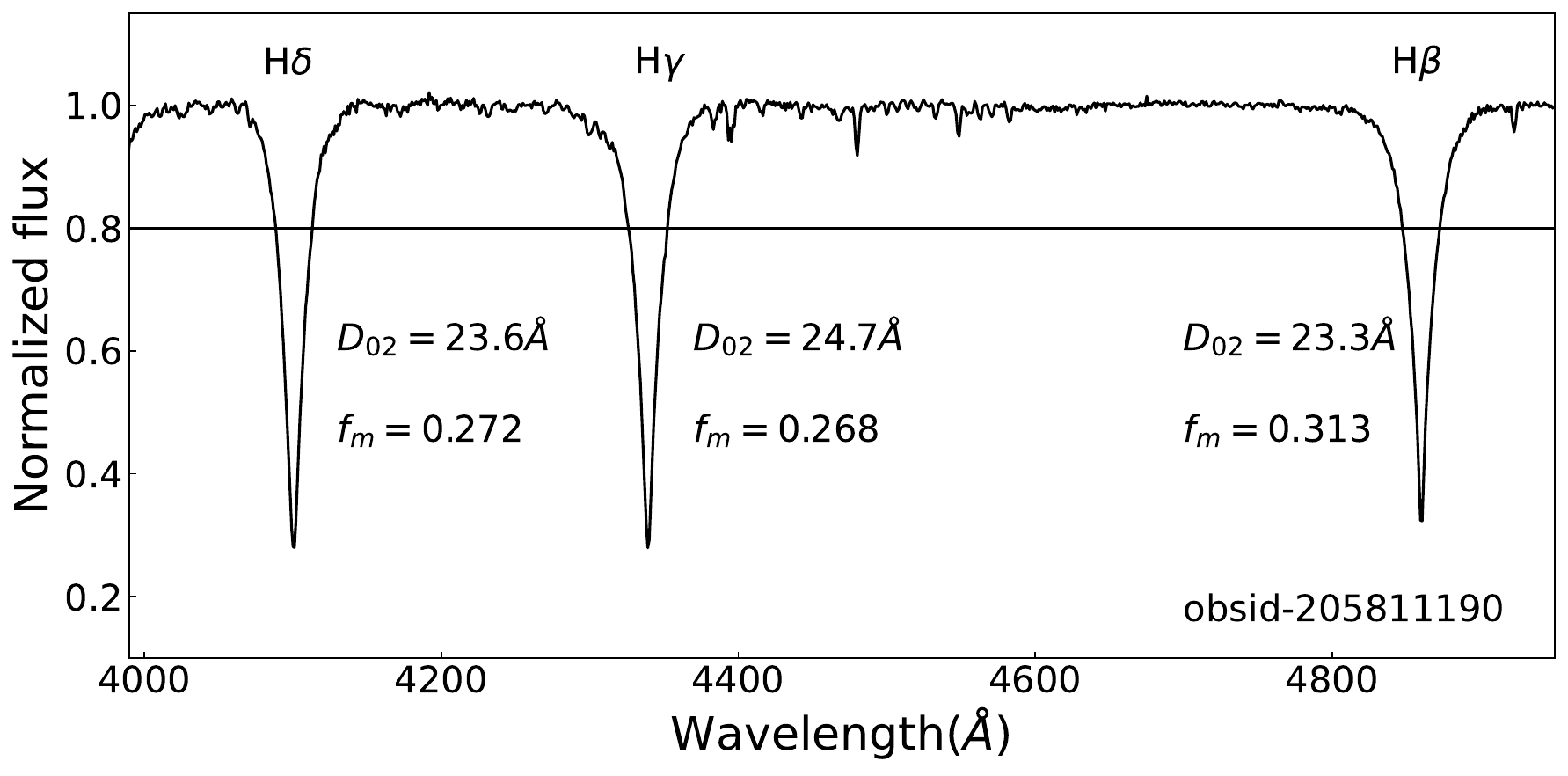}
	\caption{The spectrum of a typical BHB star with high S/N (top) and the \hdelta, \hgamma\ and $\rm{H}\beta$ region of the same star in its normalized spectrum (bottom). We also show the (\ffm, \ddo) parameters of the above three Balmer lines, which are used to select this sample spectrum are labeled. \label{fig:hgamma-hdelta}}
\end{figure}

\begin{figure}
	\epsscale{1.2}
	\plotone{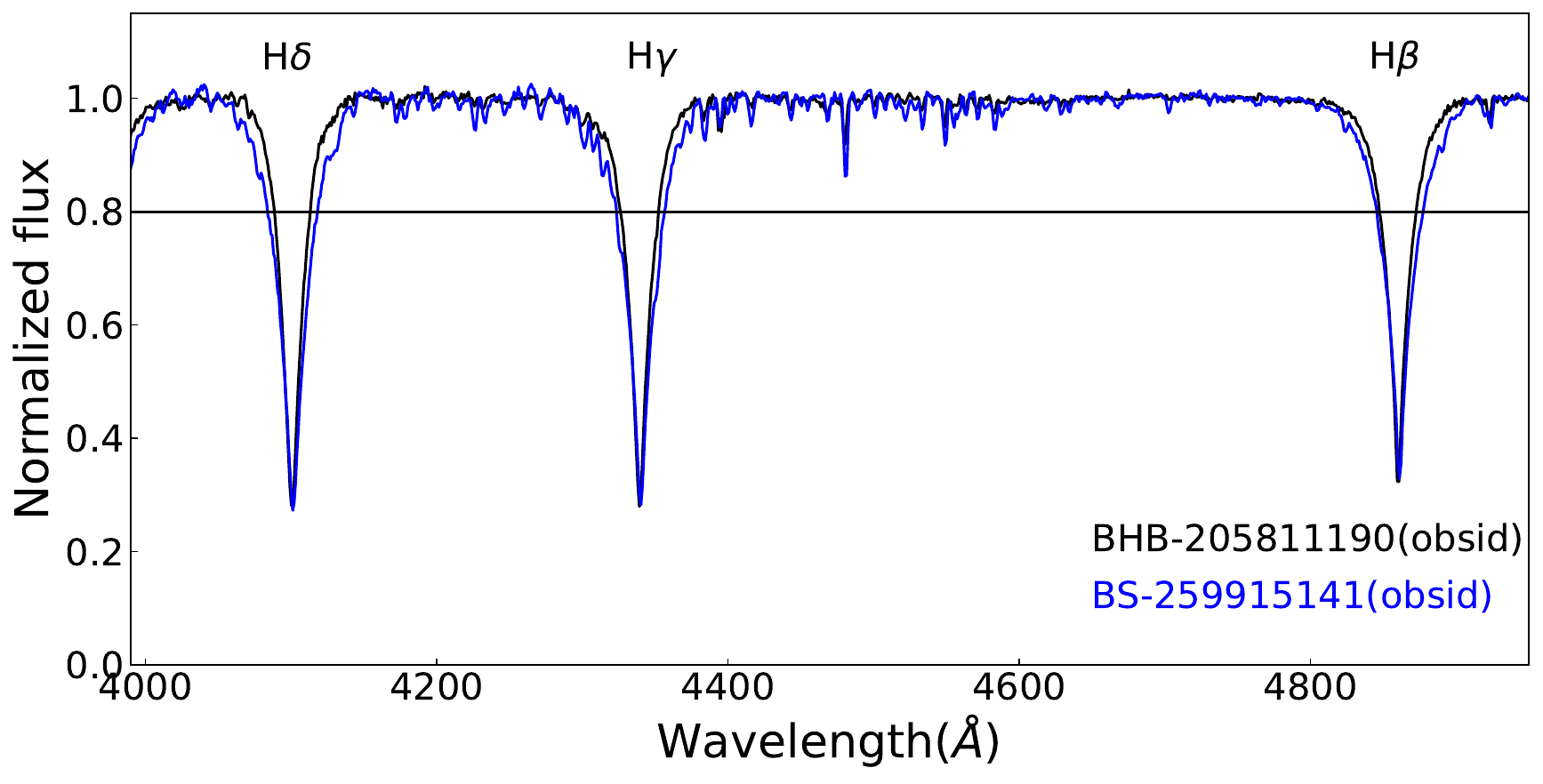}
	\caption{The \hdelta, \hgamma\ and \hbeta\ region in their normalized spectrum of a BHB star (solid line) and a BS star (dotted line) with similar effective temperatures. It is clear that the Balmer lines of the BHB star are narrower than those of the BS star at 20\% below the local continuum, which is mainly due to the lower surface gravity of BHB stars.  \label{fig:BHB-B}}
\end{figure}

 
A typical BHB spectrum is illustrated in Figure~\ref{fig:hgamma-hdelta}. The Balmer lines are strong and sharp and their line profiles are sensitive to both surface gravity and temperature. Usually, BHB stars have lower surface gravities than BS stars (see Figure~\ref{fig:BHB-B}), A- and some late B-type MS, and higher temperatures than old halo/thick disk MS stars. This has been used to distinguish BHB stars from BS, old MS stars, and 1,170 and 2,401 high Galactic latitude BHB stars are selected by \citet{Sirko2004} and \citet{Xue2008}, respectively. The way to do that is to combine two independent methods, i.e., the $D_{0.2}$ method \citep{Pier1983,Sommer1986,Arnold1992,Flynn1988,Kinman1994,Wilhelm1999a} and the scale width-shape method \citep{Clewley2002}, which will be described in details below.


For the \ddo\ method, the BHB stars can be distinguished from the BS stars with a similar temperature by measuring the \ddo\ values, i.e., the width of the Balmer line at 20\% under the normalized continuum \citep{Yanny2000}. A comparison of the Balmer line profiles for a BHB star and a BS star with similar temperature from the LAMOST DR5 is illustrated in Figure~\ref{fig:BHB-B}. In fact, the A-type MS stars can also be distinguished from the BHB stars with similar temperatures by the \ddo\ values, which have similar line profiles with the BS stars (see bottom panel of Figure 8 of Yanny et al.2000). Furthermore, the cooler MS stars can be eliminated by determining the value of \ffm, i.e., the flux relative to the normalized continuum at the line core \citep{Sirko2004, Xue2008}, as they have shallower Balmer lines.

\begin{figure*}
	\epsscale{1.2}
	\plotone{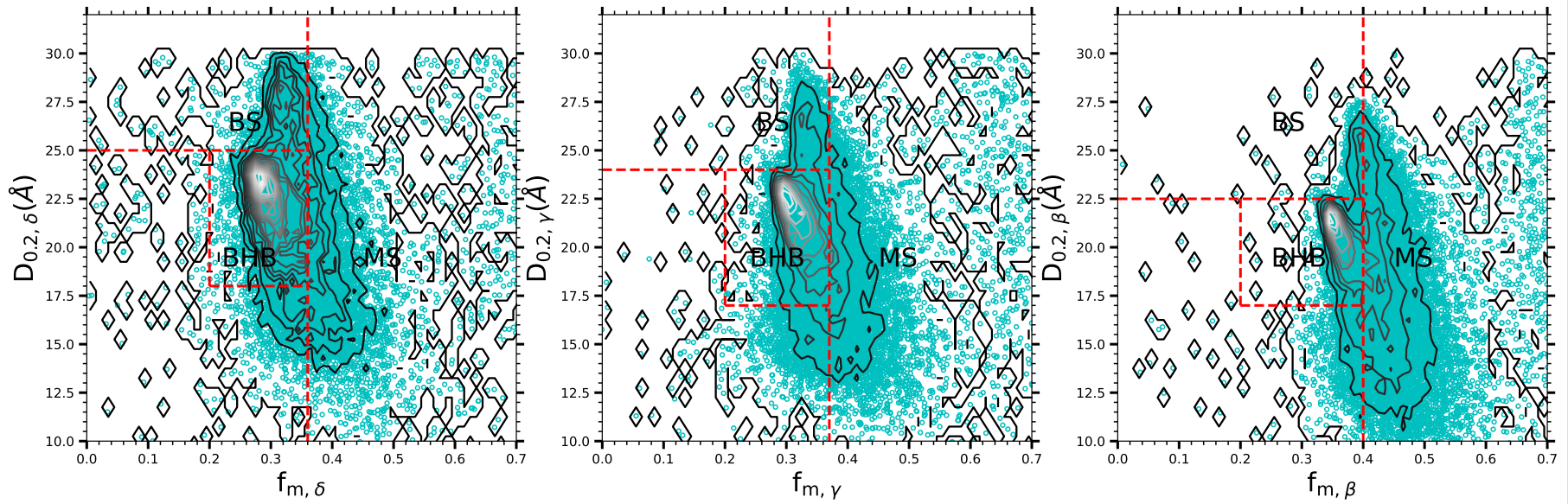}
	\caption{Distributions of the 23,528+552 BHB candidates (the contours) on the \ddo\ vs. \ffm\ plane from left to right for the \hdelta, \hgamma\  and \hbeta\ lines, respectively. The red box represents the criteria to further select BHB stars (see in details the text). \label{fig:d02-fm}}
\end{figure*}

\begin{figure*}
	\epsscale{1.15}
	\plotone{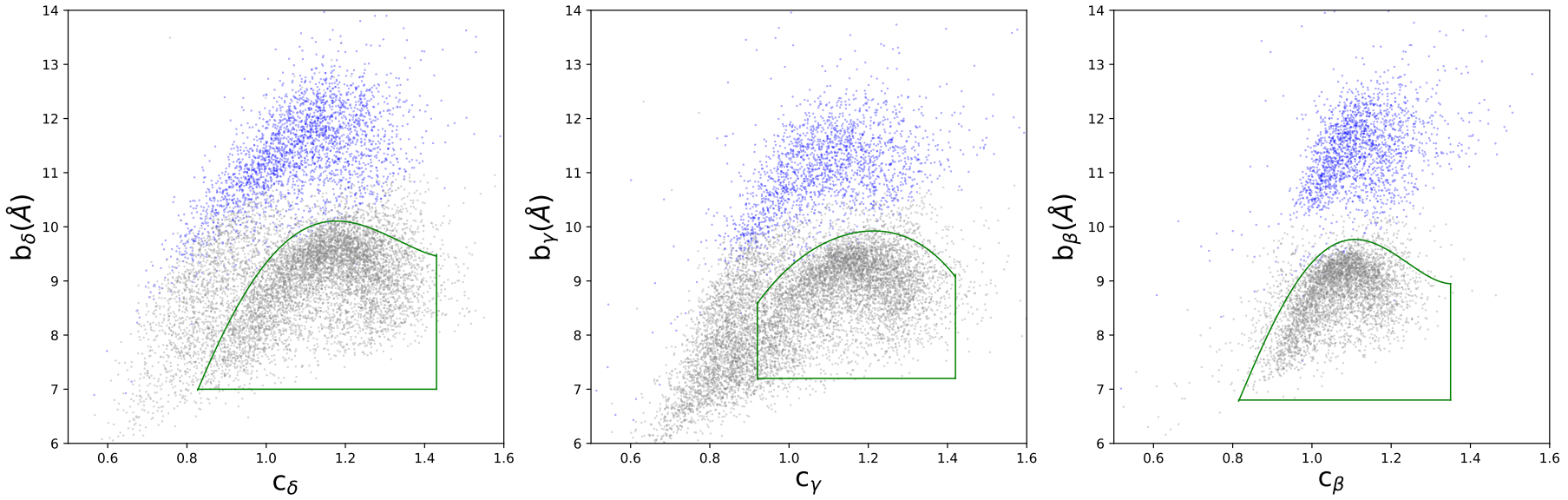}
	\caption{{Distributions of 7,955+143, 8,955+163, and 4,283+23 BHB candidates (gray dots) and BS (blue dots) from left to right for \hdelta, \hgamma\ and \hbeta\ in the $b$ versus $c$ plane.  $b$ is an effective indicator of the surface gravity and $c$ represents the temperature. The BHB candidates and BS selected from Figure~\ref{fig:d02-fm}. There is a clear gap between them. The enclosed green boxes represent the criteria for selecting BHB stars}. \label{fig:b-c}}
\end{figure*}

The scale width-shape method \citep{Clewley2002} is based on \citet{Sersic1968}  to fit the Balmer lines. \citet{Sersic1968} argues that the strong Balmer lines in the spectra of hot stars, e.g., B- and A-type stars, are no longer a Gaussian profile, but a S\'ersic profile
\begin{equation}\label{eq:sersic}
y=1.0-a\,\rm{exp}\left[-\left(\frac{|\lambda-\lambda_0|}{b}\right)^c\right].
\end{equation}  
where $y$ is the normalized flux, $\lambda$ is the wavelength, and $\lambda_0$ is the nominal wavelength of a fixed Balmer line. Based on the fitting of the Balmer line profile using equation~(\ref{eq:sersic}), \citet{Clewley2002} has proposed the scale width-shape method to identify BHB stars. Because the small uncertainties in corrections for radial velocity and normalization of spectral flux are unavoidable,  \citet{Xue2008} uses five free parameters: $a, b, c$, $\lambda_0$, and $n$ to fit the S\'ersic profile of Balmer lines after normalizing the spectrum, which is 
\begin{equation}\label{eq:sersic1}
y=n-a\,\rm{exp}\left[-\left(\frac{|\lambda-\lambda_0|}{b}\right)^c\right].
\end{equation}  


During the identification process, \citet{Xue2008} only adopts the \hgamma\ line, while \citet{Sirko2004} uses both \hdelta\ and \hgamma\ lines. In order to select more precise and more BHB stars, we calculate the five parameters, i.e. $n, a, b, c, \lambda_0$ of three Ballmer lines \hdelta, \hgamma\, and \hbeta\ of the 23,528+552 BHB candidates through fitting their line profiles using equation~(\ref{eq:sersic1}). 

\subsection{Identification of BHB stars with the \ddo\ method and the scale width-shape method}

 Firstly we map the 23,528+552 BHB candidates on the \ddo\ vs. \ffm\ plane from left to right for the \hdelta, \hgamma\, and \hbeta\ lines, respectively (see Figure~\ref{fig:d02-fm}). Compared to Figure 4 of \citet{Xue2008}, the areas of the location of BHB, BS, and MS stars are labeled, respectively. The contours show that the BHB stars accumulate on the left middle area of each panel of Figure~\ref{fig:d02-fm}. 
The concentration of stars centered at $(f_{m,\delta}, D_{0.2,\delta})=(0.31, 22.00\,\mathrm{\AA})$, $(f_{m,\gamma}, D_{0.2,\gamma})=(0.32, 0.65\,\mathrm{\AA})$, and $(f_{m,\beta}, D_{0.2,\beta})=(0.37, 20.05\,\mathrm{\AA})$ represents the BHB stars and the stars with larger \ddo\ are BS stars and the stars with smaller \ddo\ and \ffm\ are cooler giants, but with larger \ffm\ are late A-type MS stars. For the concentration site of BHB stars, our results have slightly different with the results from the literature, i.e., at ($f_{m,\delta}, D_{0.2,\delta}) = (0.30, 26\,\mathrm{\AA})$ presented by \citet{Sirko2004} and (0.23, 25\,\AA) by \citet{Xue2008}, which should be due to the difference between the resolution of the spectra used and the fitting program adopted. 
Based on the criteria of \citet{Xue2008}, we select BHB stars with the following empirical criteria:
\begin{equation}
\begin{split}
\label{d02}
&18\leq\ {D}_{0.2,\delta}\leq25\,\mathrm{\AA}, 0.2\leq f_{m,\delta}\leq0.36;\\
&17\leq {D}_{0.2,\gamma}\leq24\,\mathrm{\AA}, 0.2\leq f_{m,\gamma}\leq0.37;\\
&17\leq\ {D}_{0.2,\beta}\leq22.5\,\mathrm{\AA}, 0.2\leq f_{m,\beta}\leq0.4.
\end{split}    
\end{equation}

After applying these cuts, we obtain 7,955+143, 8,955+163, and 4,283+23 BHB candidates for the criterion of \hdelta, \hgamma\, and \hbeta, respectively. 

Then we use the scale width-shape method to further remove the contamination from the BS and cooler MS stars. Here $b$ is an effective indicator of the surface gravity, i.e., higher $b$ corresponds to larger surface gravity at fixed $c$, and $c$ represents the temperature, in which the cooler stars with lower $c$ at fixed $b$.

Figure~\ref{fig:b-c} shows the distributions of 7,955+143, 8,955+163, and 4,283+23 BHB candidates in the $b$ versus $c$ plane from left to right for \hdelta, \hgamma\ and \hbeta, respectively.
We also select 3,367, 2,417, and 2,239 BS stars for \hdelta, \hgamma\, and \hbeta, respectively, based on their distributions, and plot them in Figure~\ref{fig:b-c} for comparison.

Figure~\ref{fig:b-c} displays bimodal distributions up and down when $c$ is larger than 0.8, 0.9, and 0.8 from left to right in the three panels. $b$ is an effective indicator of the surface gravity and $c$ represents the temperature. The reason is that the BS stars are larger than those of BHB stars with a similar temperature.  It helps us to remove further the BS stars that still remain in our BHB candidates. The stars located in the left lower region with small $c$ values are the cooler MS stars. Based on the distributions of BHB stars in Figure~\ref{fig:b-c}, we cleanly select BHB stars with the following criteria:
\begin{equation}
\begin{split}
\label{b-c}
&b_{\delta}\leq 45.52c_{\delta}^4-184.6c_{\delta}^3+252.9c_{\delta}^2-124.9c_{\delta}+20.45,\\
&b_{\delta}\geq 7, 0.83\leq c_{\delta}\leq 1.43;\\
&b_{\gamma}\leq -38.92c_{\gamma}^4+177c_{\gamma}^3-316.3c_{\gamma}^2+263.7c_{\gamma}-76.31,\\
&b_{\gamma}\geq 7.2, 0.92\leq\rm{c}_{\gamma}\leq 1.42;\\ 
&b_{\beta}\leq 135.1c_{\beta}^4-553.4c_{\beta}^3+811.3c_{\beta}^2-494.8c_{\beta}+111.1,\\
&b_{\beta}\geq 6.8, 0.82\leq c_{\beta}\leq 1.35.
\end{split}
\end{equation}

\begin{table*}
\footnotesize
\centering
    \caption{BHB stars identified in LAMOST DR5. }\label{Table 2}
    \begin{tabular}{ccccccccccc}
    \hline
    obsid &        RA &        Dec &     S/N &     $T_{\rm eff}$ &   log\,$g$ &   [Fe/H] & $BP-RP$&  E(B-V)  \\
     &($^{\circ}$)  & ($^{\circ}$) & ($g$-band)&(K) &  && (Gaia EDR3)&  \\
    \hline
    66604134 &  0.040234  &  11.892757 &  62.24 &  10228.93  &  3.13&  -1.55 &  0.122512& 0.09693\\
83814139 &  0.127198  &  32.134235 &  159.08&  8160.17  &  3.20&  -1.71 &  0.234762&0.04944 \\
250504090&	0.184691&	37.842845&	181.02&	7369.15&	3.07&	-0.84&	0.516507&	0.09709\\
87514160 &  0.254373  &  37.285919 &  21.17 &  10198.65  &  3.71 &  -1.82  &  0.1724& 0.09495  \\
14812243 &  0.2615792 &  33.3272278&  36.63 &  9201.36  &  3.09&  -1.56  &  0.124019&0.04829 \\
8614049  &  0.266271  &  5.169305  &  35.08 &  8370.94  &  3.63&  -1.38  &  0.264092&0.03958 \\
55001071 &  0.371612  &  16.030824 &  16.46 &  7534.25 &  2.91&  -1.43  &  0.334815&0.03706 \\
385811162&  0.387119  &  22.974005 &  28.11 &  8849.76  &  3.32 &  -1.94  &  0.213063&0.07867 \\
281910151&	0.5106794&	35.836533&	35.65&	8349.24&	4.18&	-1.13&	0.384477&	0.09779\\
250509157&  0.5265322 &  39.183206 &  127.71&  10211.44 &  3.80&  -1.00  &  0.107694&0.11994 \\ 
281903228&  0.5786953 &  36.286141 &  41.57 &  9351.78  &  3.26&  -1.23  &  0.17907&0.10611  \\
496108188&  0.5861244 &  6.7466514 &  83.7  &  10279.50 &  3.70&  -1.52  &  0.099919&0.04314 \\
66111030 &  0.594877  &  4.922824  &  30.42 &  7656.46 &  3.08&  -1.71  &  0.263664&0.02419 \\
492104104&  0.606479  &  25.600805 &  27.35 &  9294.287  &  2.95 &  -1.65  &  0.119936&0.03575 \\
385804175&	0.730444&	20.560251&	67.13&	10263.31&	3.53&	-0.85&	0.072537&	0.0538\\
181316089&  0.874317  &  5.999652  &  67.25 &  9016.02 &  2.97&  -2.20  &  0.155702& 0.05402\\
66608022 &  0.877783  &  11.732231 &  55.68 &  8666.09  &  3.15&  -1.64  &  0.149648&0.06578 \\
472911205&	0.9361385&	19.7528826&	18.14&	9634.15&	3.80&	-0.98&	0.155636&	0.04098\\
369207168&  0.945362  &  0.635879  &  30.28 &  9026.94   &  3.52&  -1.94   &  0.159239 &0.02984\\

    ...	&...	&...	&...  &...	&...	&...	&... \\

    \hline& 
    \end{tabular}
    \tablecomments{The first column represents the object names. and the next two give the object astrometry (RA, Dec). The signal to noise at the g band is in the next column. The next three columns contain the atmospheric parameters estimated by \citet{Xiang2022}. The next column is the color ($BP-RP$ ) obtained from Gaia EDR3. The last column is E(B-V) accepted from \citet{Schlegel1998}}
    
    \end{table*}

After selecting BHB stars with the above method, we obtain 5,578+ 85 BHB spectra by merging the groups of \hdelta, \hgamma, and \hbeta. After removing duplicate sources, we finally obtain 5,355 BHB stars and 81 BHB candidates, which are listed in Table 1. It should be noted again that there may be non-BHB stars with [Fe/H]\,$\geq$ -1 in the 81 BHB candidates, which have similar Balmer line profiles with the BHB stars. 


\subsection{Completeness}

\begin{figure}
	\epsscale{1.0}
	\plotone{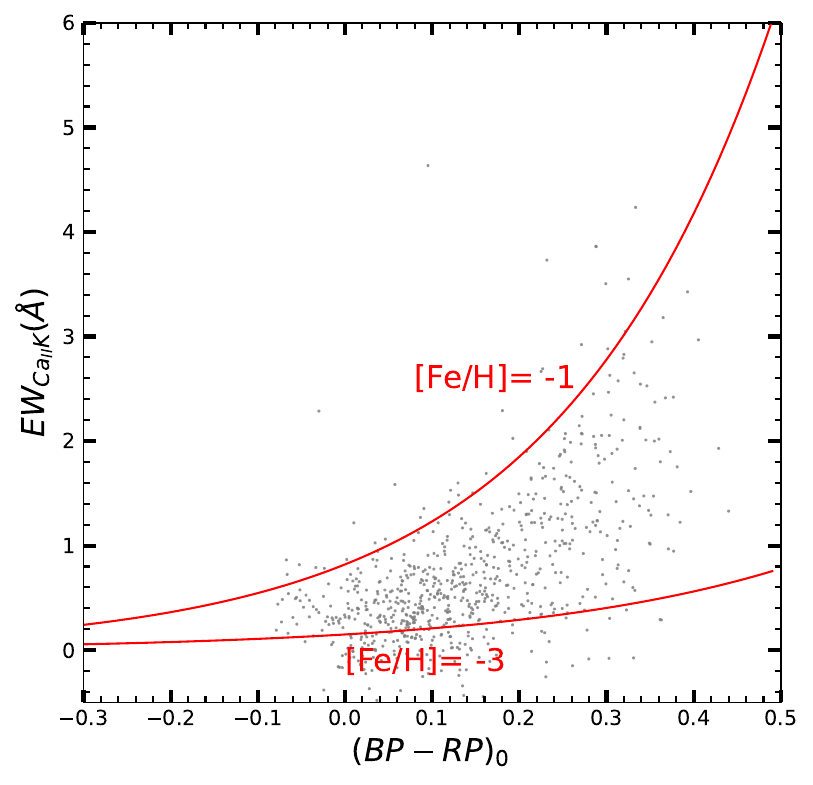}
    \plotone{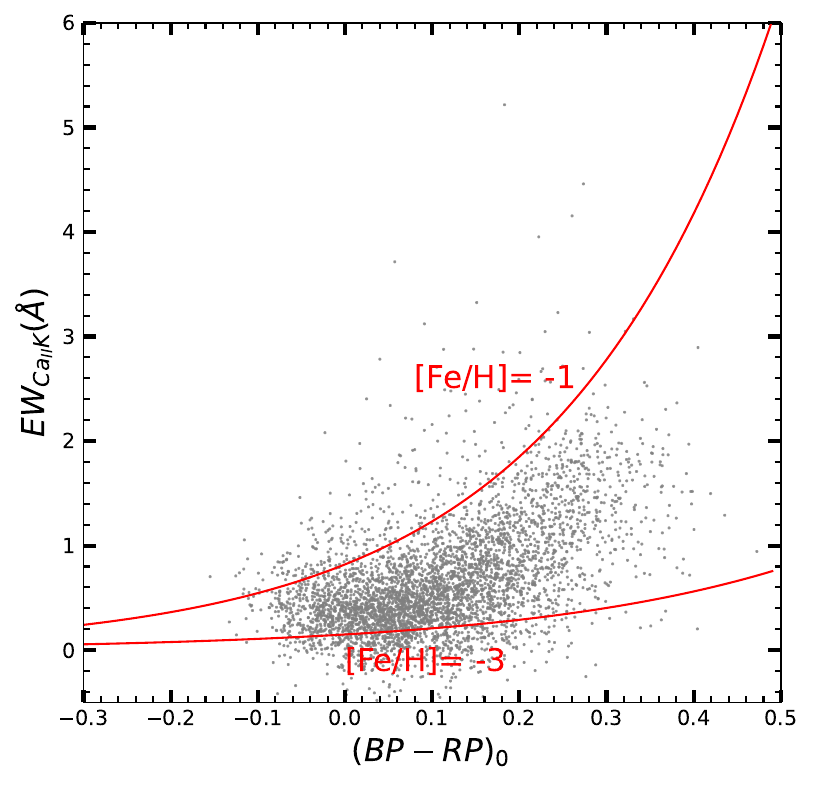}
	\caption{The distributions of 765 \citep{Xue2011} (the top panel) and 5,198 BHB stars \citet{Vickers2021} (the bottom panel) on the $\mathrm{EW_{Ca\,II\,K}}$ versus $(BP-RP)_0$ plane. These BHB stars have  (S/N)$_g\le10$ and $|Glat|>20^{\circ}$ and are cut by line indices. The red solid lines are the same in Figure~\ref{fig:Cak-feh-1}.}\label{fig:wk-sh-sdss}
\end{figure}



The BHB star catalogs of \citet{Xue2011} and \citet{Vickers2021} are used as reference catalogs to verify the completeness and reliability of our sample by using the selection method adopted by this work. \citet{Xue2011} obtained 4,985 BHB stars from SDSS DR8 by using the color cuts and the cuts based on fitting the profile of Balmer line \hgamma. The BHB catalog constructed by \citet{Vickers2021} contains 13,693 spectra of 11,046 BHB stars selected through a machine-learning algorithm based on LAMOST DR5. 
 
We first cross-match the two catalogs with that of LAMOST DR5 and obtain 1,055 and 12,075 spectra, respectively. Secondly, we eliminate the spectra with (S/N)$_g\le10$ and $|Glat|\leq20^{\circ}$, then 779 and 5,310 BHB stars left, respectively. 

Through the cuts of equation (\ref{eq:line-ind}), 765 and 5,198 BHB stars are obtained from the 779 and 5,310 samples, respectively.
Figure~\ref{fig:wk-sh-sdss} shows the distributions of 765 \citep{Xue2011} (the top panel) and 5,198 BHB stars \citep{Vickers2021} (the bottom panel) on the $\mathrm{EW_{Ca\,II\,K}}$ versus $(BP-RP)_0$ plane. Applying the cuts of metallicity, i.e., $\rm{[Fe/H]}\le-1$, 730 and 4,826 BHB stars are obtained from the 765 and 5,198 samples, respectively. 
Then applying the cuts of equations~(\ref{d02}) and( \ref{b-c}), 492, 426, and 264 BHB stars are left based on the fitting results of \hdelta, \hgamma\ and \hbeta\ lines for the \citet{Xue2011} sample, and 3,866, 3,633 and 2,616 BHB stars left for the \citet{Vickers2021} sample, respectively.  
After merging and removing duplicated sources, 501 and 3,963 BHB stars are finally obtained from the 779 and 5,310 samples, respectively. It indicates that about 64.3\% of BHB stars of \citet{Xue2011} and 74.6\% of BHB stars of \citet{Vickers2021} are selected using the criteria adopted by this work. If the metal-rich stars were not excluded, the fractions should be slightly larger. 

\section{Discussion}

\subsection{Atmospheric Parameters of BHB stars }

\begin{figure*}
	\epsscale{1}
	\includegraphics[width=1\textwidth]{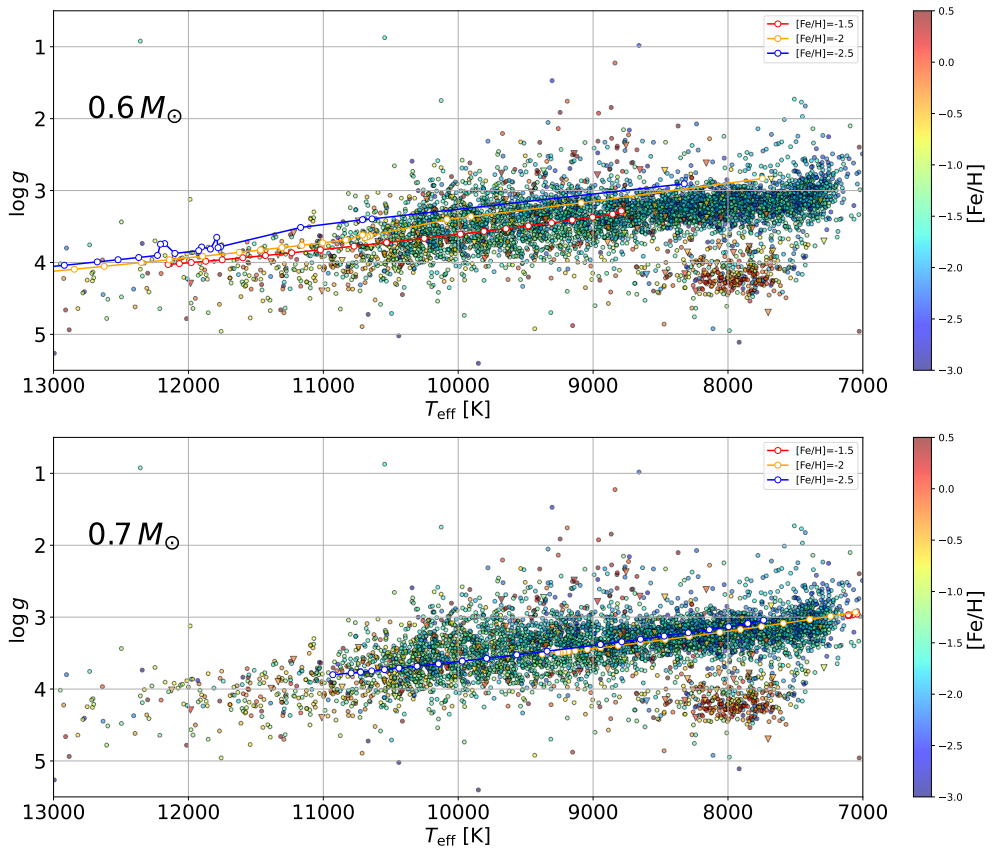}
	\caption{The distribution of our 5,330 + 81 BHB sample (unfilled circle) on the log\,$g$ vs. $T_{\rm eff}$ plane, which is adapted from \citet{Xiang2022}. The colorbar represents the metallicities of the BHB stars. The unfilled circle lines represent the evolutionary tracks for BHB stars with the mass 0.6\,$M_\odot$ (the upper panel), 0.7\,$M_\odot$ (the bottom panel), and [Fe/H]$= -1.5$ (red), $-2$ (orange), $-2.5$ (blue), respectively.  \label{fig:6808track}}
\end{figure*}

\citet{Xiang2022} has estimated the atmospheric parameters of OBA-type stars from LAMOST DR6 through HotPayne Methods. We cross-match our BHB stars and \citet{Xue2011} sample with it and get the atmospheric parameters for 5,330 + 81 and 1,030 BHB stars, respectively. About 25 of our BHB stars have no atmospheric parameters, which are marked in Table \ref{Table 2}. We show the distribution of these BHB stars on the log\,$g$ vs. $T_{\rm eff}$ plane in Figure~\ref{fig:6808track}, and the colorbar represents the metallicities. For comparison, the evolutionary tracks of BHB stars with the masses 0.6, 0.7 ${\rm M}_{\odot }$, and the metallicities [Fe/H]$= -1.5, -2, -2.5$ are also presented in this figure, which is calculated by \citet{Fu2018}.

\citet{Catelan2009} pointed out that the effective temperatures of the BHB stars generally range from 7,000 to 20,000\,K, and from 2.5 to 4.5,  for the surface gravities, respectively. From Figure~\ref{fig:6808track}, we can see that the variation trend of the log\,$g$ with $T_{\rm eff}$ for most of the BHB stars is well consistent with the theoretical tracks. In addition, most of our BHB stars are located in the range of around 7,200-11,000\,K for the $T_{\rm eff}$ and 2.8-4.2,  for the log\,$g$, which confirms the fact that our sample is mainly the A-type BHB stars, and also old and low mass objects. In other words, it indicates that most of our selected BHB stars are reliable. It is noted that there are about 350 BHB stars with log\,$g$ about 4.5,  which is larger than most of the BHB stars with similar $T_{\rm eff}$ around 7,000-8,000\,K. Moreover, \citet{Xiang2022} shows that most of these 350 BHB stars also have large metallicities, even close to 0.5 of [Fe/H]. The possible reasons are that our samples may still have some contamination from the metal-rich dwarfs or some bias in the estimated atmospheric parameters for these BHB stars. In fact, nearly half of the spectra of these larger log\,$g$ stars have  (S/N)$_g$ $<$ 30, which may affect the estimation of their atmospheric parameters.
 
\begin{figure*}
	\epsscale{1.0}
	\plotone{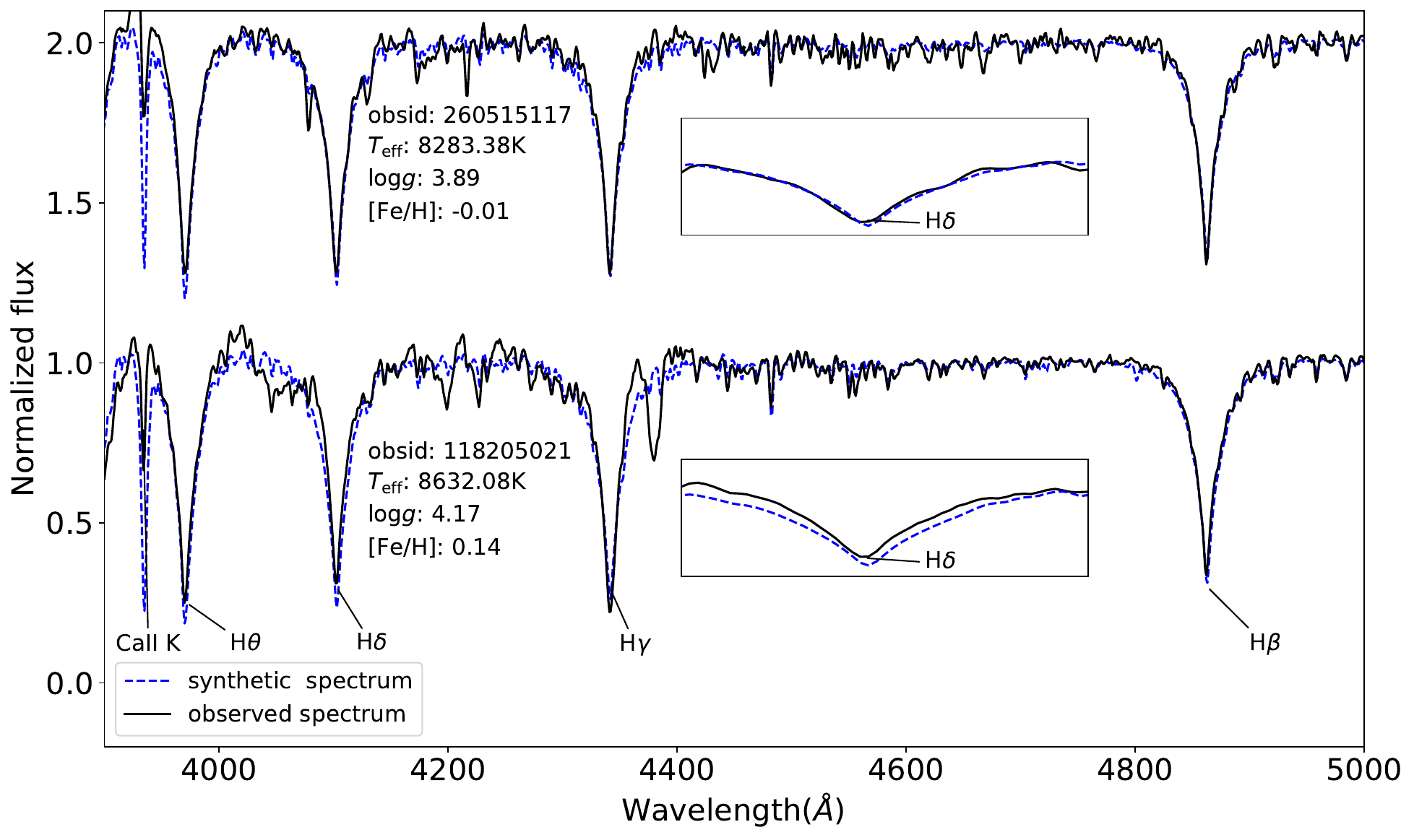}
	\caption { The normalized spectrum of two stars with larger log\,$g$. The black lines show the normalized spectrum of the star. The blue dashed line shows the synthetic spectra.\label{fig:out}}
\end{figure*}

To further check the reliability of our BHB sample, we select two spectra of the above 350 BHB candidates with (S/N)$_g =$ 220.05 and 237.13, respectively. Their corresponding synthetic spectra \citep{Allende2018} are calculated using the atmospheric parameters estimated by \citet{Xiang2022}, which are also shown in Figure~\ref{fig:out} for comparison. From Figure~\ref{fig:out}, we can see that the synthetic spectra have strong \ion{Ca}{2}\,K absorption and wider H$\delta$ line profile than those in the observed stellar spectra. This means that these stars should have lower metallicity than the synthetic counterpart or they may suffer from mismatched temperature and surface gravity.


\begin{figure}
	\epsscale{1.2}
	\plotone{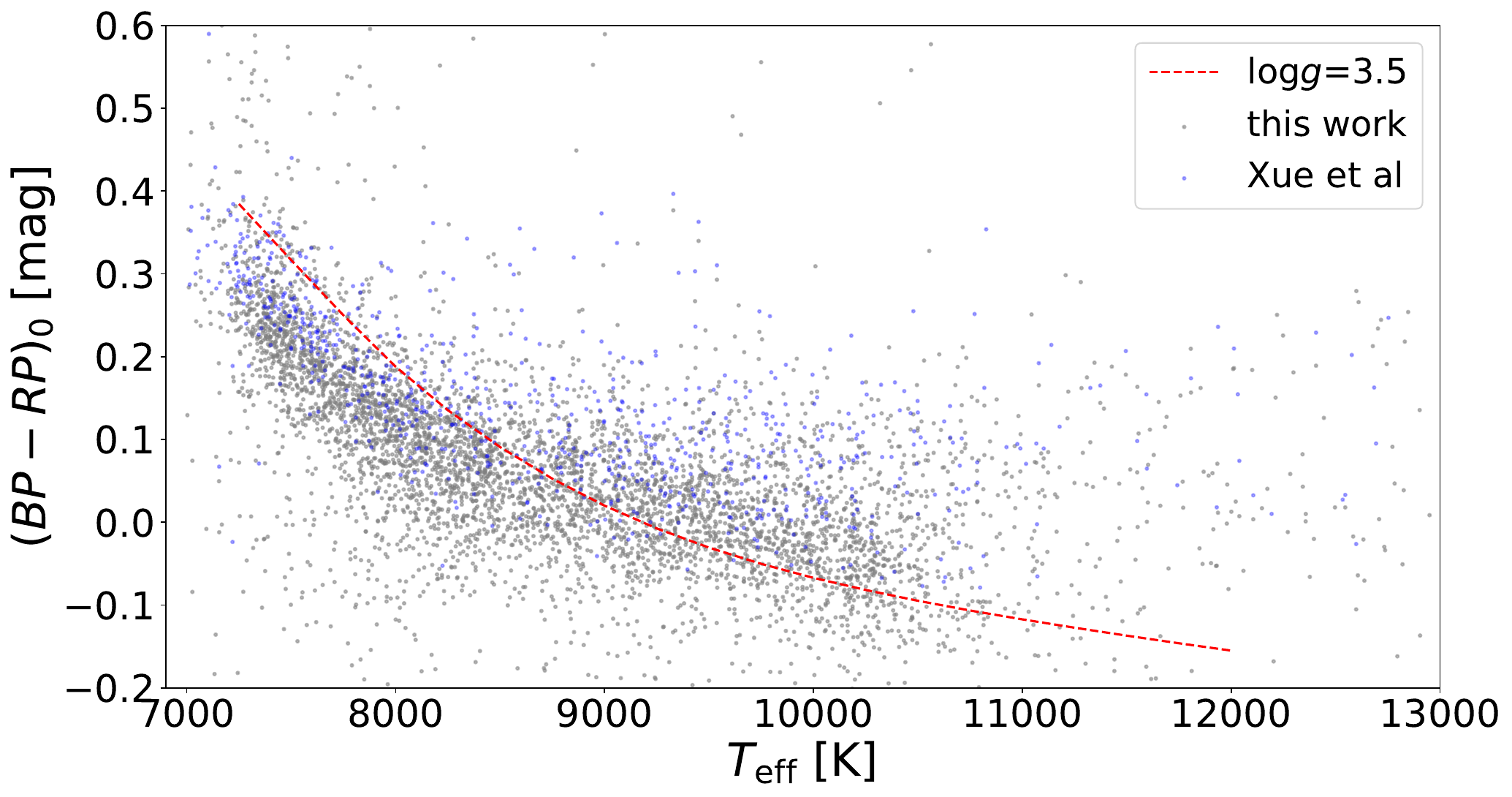}
	\caption {$(BP-RP)_0$ - $T_{\rm eff}$ diagram for our BHB stars (gray filled circle) and \citet{Xue2011} BHB stars (blue filled circle), $(BP-RP)_0$ is corrected for extinction \citep{Schlegel1998}. $T_{\rm eff}$ is estimated by \citet{Xiang2022}. The red dotted line represents theoretical results calculated by the PARSEC database with [Fe/H]$= -1.$5 and log\,$g = 3.5$. \citep{Chen2019}. \label{fig:bp-t}}
\end{figure}

We display $(BP-RP)_0$  of our BHB stars as a function of $T_{\rm eff}$ estimated by \citet{Xiang2022} in Figure~\ref{fig:bp-t}, and \citet{Xue2011} sample is also presented for comparison. The prediction of the  PARSEC  for $(BP-RP)_0$ vs. $T_{\rm eff}$ is also presented \citep{Chen2019, Paxton2011, Choiet2016}. For most of our and \citet{Xue2011} BHB stars, we can see that their $(BP-RP)_0$ vs. $T_{\rm eff}$ follow a similar anti-correlation, which is consistent with the prediction of the PARSEC stellar model with [Fe/H]$= -1.5$ and log\,$g = 3.5$ \citep{Chen2019}, although there are a few BHB stars in Figure~\ref{fig:bp-t} whose $(BP-RP)_0$ colors and $T_{\rm eff}$ are inconsistent.

To sum up, most of these stars are BHB stars, and a few contaminated stars which may be due to the bias in the parameters estimation. It requires further verification using high-resolution spectra. 


\subsection{Comparison with previous work}

\citet{Vickers2021} claimed that they obtained 13,693 spectra of 11,045 BHB stars also from LAMOST DR5, which is a little more than double of our BHB stars. They gather a stellar sample with colors indicating $T_{\rm eff}>$ 7,000\,K, which is observed by both LAMOST and Gaia. Then they use the LAMOST spectral data with labeled stellar classifications and metallicities which are from -3 to -0.31 to train a machine learning algorithm. The machine finally produces a catalog of BHB stars with their metallicity information. 
The Balmer-line profiles in the spectra of the BHB stars and the metal-rich A-type main sequence stars are very similar, which is not easy to distinguish. We first remove the metal-rich disk stars and the stellar spectra with low signal-to-noise using the criteria $|Glat|<20^{\circ}$ and (S/N)$_g\le10$, respectively. 
After using these two criteria, only 5,310 of 11,045 BHB stars from \citet{Vickers2021} are left, which indicates that more than half of their BHB stars are metal-rich disk stars or those with (S/N)$_g\le10$. This may be due to the different selection methods.

\begin{figure}
	\epsscale{1}
    \plotone{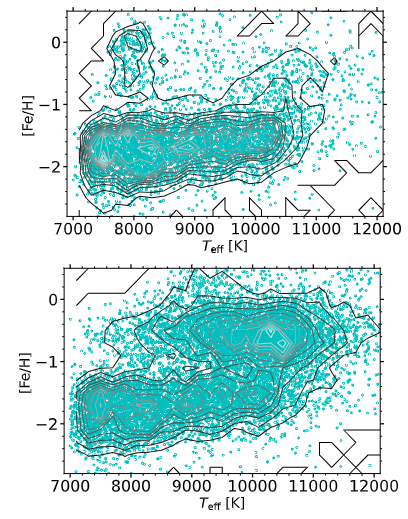}
	\caption{The distribution of our sample (top) and \citet{Vickers2021} (bottom) BHB sample in $T_{\rm eff}$ vs. [Fe/H] plane.  \label{fig:t-feh}}
\end{figure}

We compared the atmospheric parameters of $T_{\rm eff}$, log\,$g$, and [Fe/H] from \citet{Xiang2022} of BHB stars with (S/N)$_g\ge10$ in these two samples. According to the metal-poor nature ([Fe/H]$\le -1$) of BHB stars \citep{Kinman2000}, our pollution rate is only 16.6\%  significantly lower than 47.4\% of \citet{Vickers2021}. Figure~\ref{fig:t-feh} shows the $T_{\rm eff}$ vs. [Fe/H] plane for our sample (top) and that of \citet{Vickers2021} (bottom). It can be seen that the metallicity of our samples is basically less than -1 in addition to a little with [Fe/H] $\ge -1$ near 8,000\,K (discussed above) and 10,000\,K, which is due to the \ion{Ca}{2}\,K absorption lines are relatively weak at the high temperatures, and can only be roughly estimated. The sample of \citet{Vickers2021} is clearly divided into two parts by the metallicity, of which 5,172 stars have [Fe/H] larger than -1. Our samples show higher purity, which confirms the reliability of our method.

We also carried out a comparison of the effect of (S/N)$_g$ on the two samples (see Figure~\ref{fig:snr-pure}). The red solid line represents our sample, and the blue solid line represents the sample of \citet{Vickers2021}. The purity is assessed as a function of (S/N)$_g$  based on the atmospheric parameters estimated by \citet{Xiang2022}. Purity here refers to the number of the BHB sample stars with [Fe/H] $\le-1$ and a fixed (S/N)$_g$ value is divided by the total number of BHB sample stars with the same (S/N)$_g$ value. The purity of our sample is consistently surpassing 80\% with the (S/N)$_g$ increasing. However, the purity of \citet{Vickers2021} initially starts at 52.6\% when (S/N)$_g= 10$ and shows a decreasing trend with the (S/N)$_g$ increasing. At the (S/N)$_g$ of 50, the purity of \citet{Vickers2021} drops to only 39.9\%. These findings suggest that increasing the (S/N)$_g$ does not necessarily guarantee higher purity.

\begin{figure}
	\epsscale{1}
    \plotone{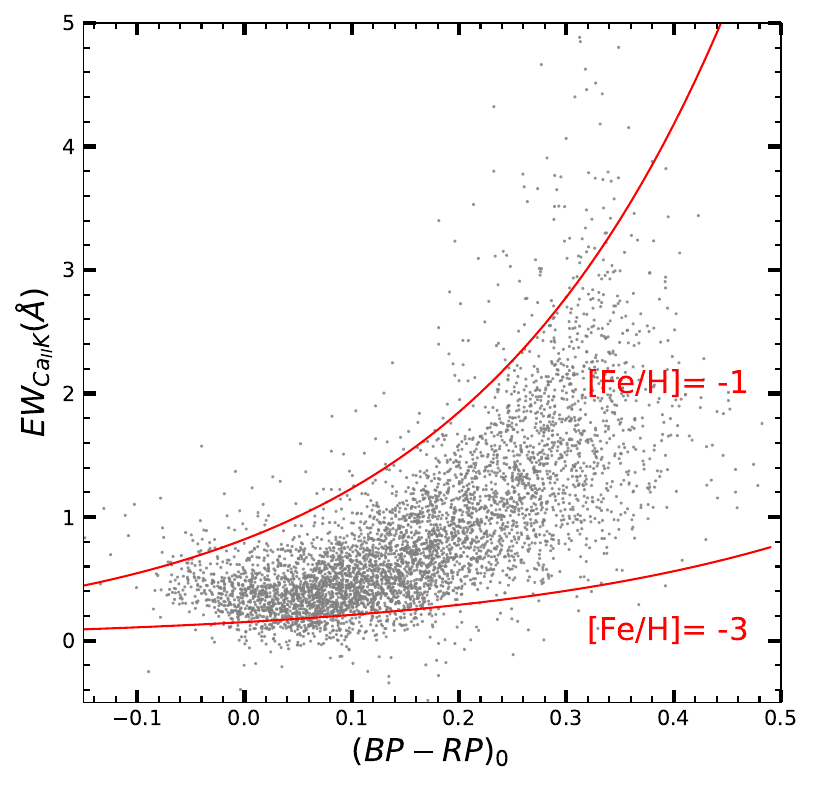}
    \plotone{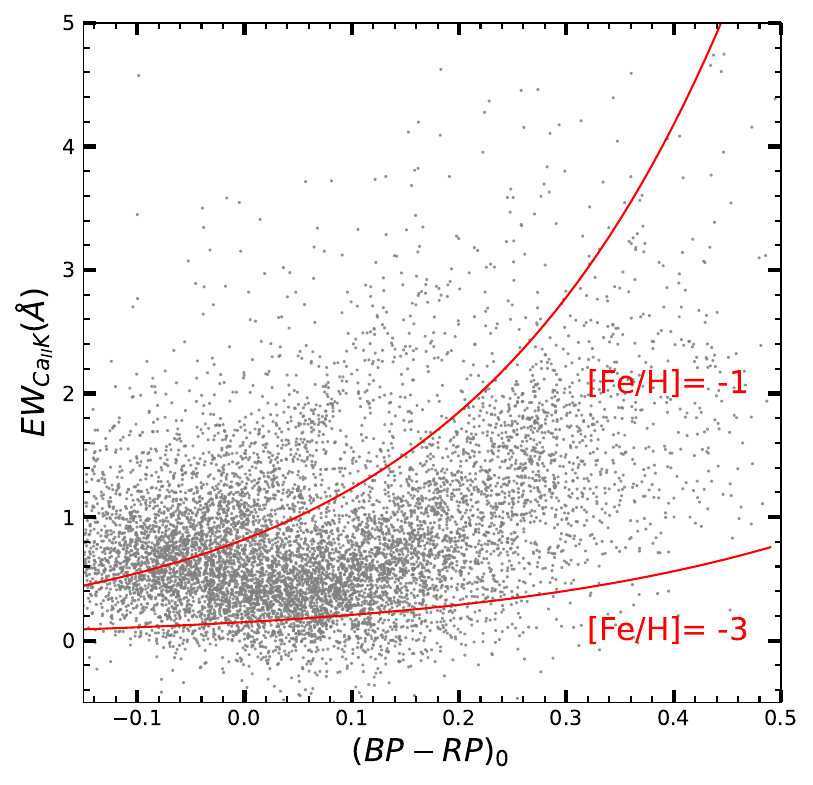}
	\caption{The distribution of \citet{Xue2011} (top) and \citet{Vickers2021} (bottom) BHB sample in $(BP-RP)_0$ vs. $\mathrm{EW}_\mathrm{Ca\,II\,K}$ plane. The red solid lines are the same in Figure~\ref{fig:Cak-feh-1}. \label{fig:sh5666-cak}}
\end{figure} 

Figure~\ref{fig:sh5666-cak} shows the $(BP-RP)_0$ vs. $\mathrm{EW}_\mathrm{Ca\,II\,K}$ for the 
BHB stars of \citet{Xue2011} and \citet{Vickers2021} 
. The $\mathrm{EW}_\mathrm{Ca\,II\,K}$ of most of the \citet{Xue2011} BHB stars range from -3 to -1, which also indicates the reliability of the method of metallicity cuts. It is noted that there are 4,367 BHB stars with [Fe/H] $>-1$ from \citet{Vickers2021} and the contamination rate is 42.5\% which is estimated according to the theoretical criteria of $\mathrm{EW}_\mathrm{Ca\,II\,K}$.

\begin{figure}
	\epsscale{1}
    \plotone{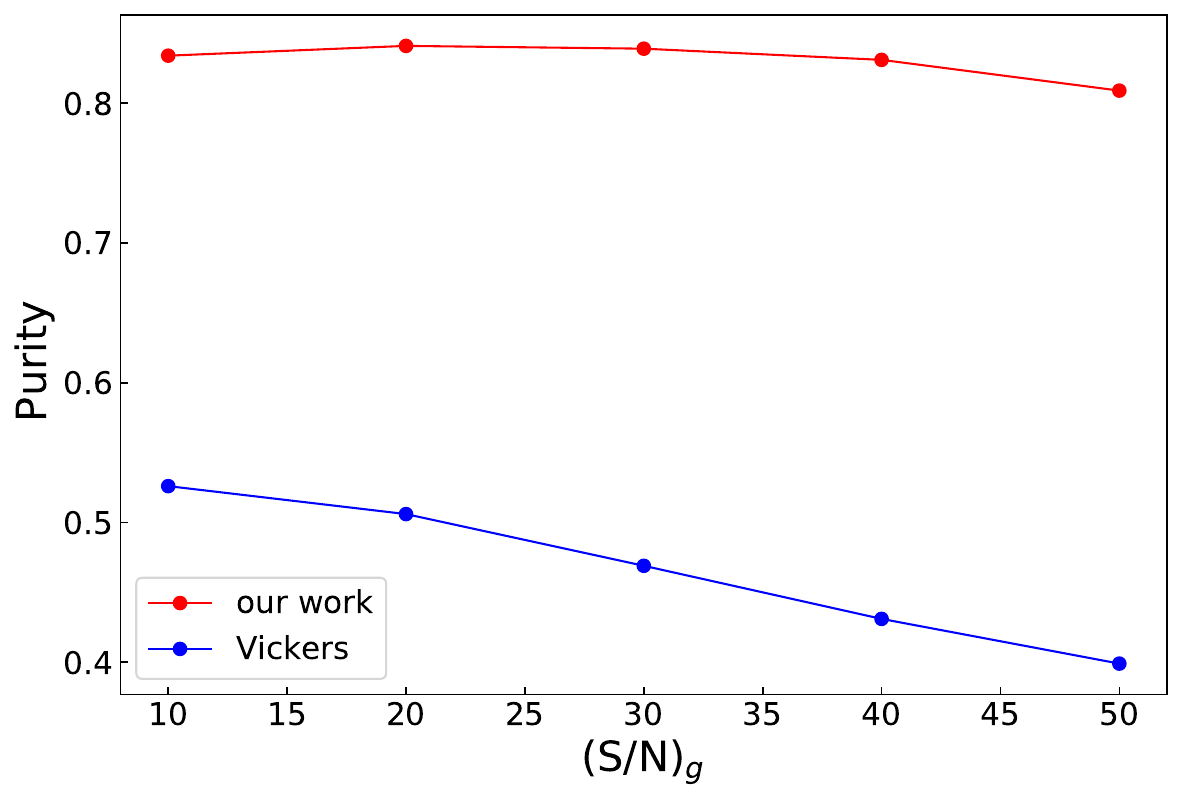}
    \caption{The purity versus  (S/N)$_g$ plots of our sample (red) and \citet{Vickers2021} (blue).\label{fig:snr-pure}}
\end{figure}

Among the 5,355+81 BHB stars in our work, only 548 stars have already been listed by \citet{Xue2011}. The main reason is that the LAMOST sources are bright, while the SDSS survey mainly observes the stars located in the Galactic halo, which is usually dim.

\subsection{Effect of (S/N)$_g$ on the selection of the BHB stars}

\begin{figure}
	\epsscale{1.1}
    \plotone{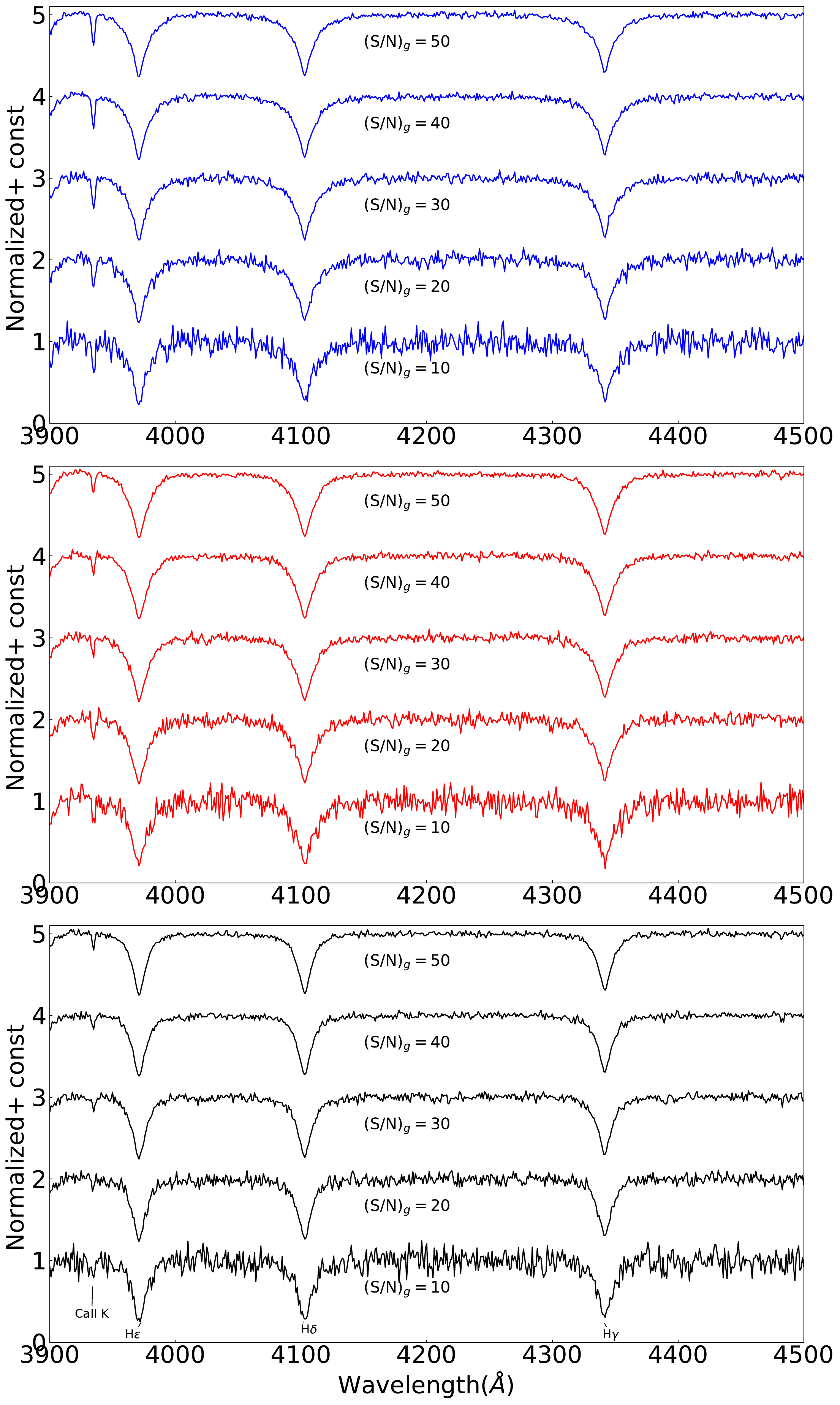}
    \caption{Synthetic spectra with log\,$g=3.5$ and [Fe/H]$\,=-1.7$, different  (S/N)$_g$ (10-50) and temperatures. From top to bottom, blue represents $T_{\rm eff}= 8,000\,K$, red represents $T_{\rm eff}= 9,000\,K$, black represents $T_{\rm eff}= 10,000\,K$.}\label{fig:snr-ll}
\end{figure}

\begin{figure}
	\epsscale{1.1}
    \plotone{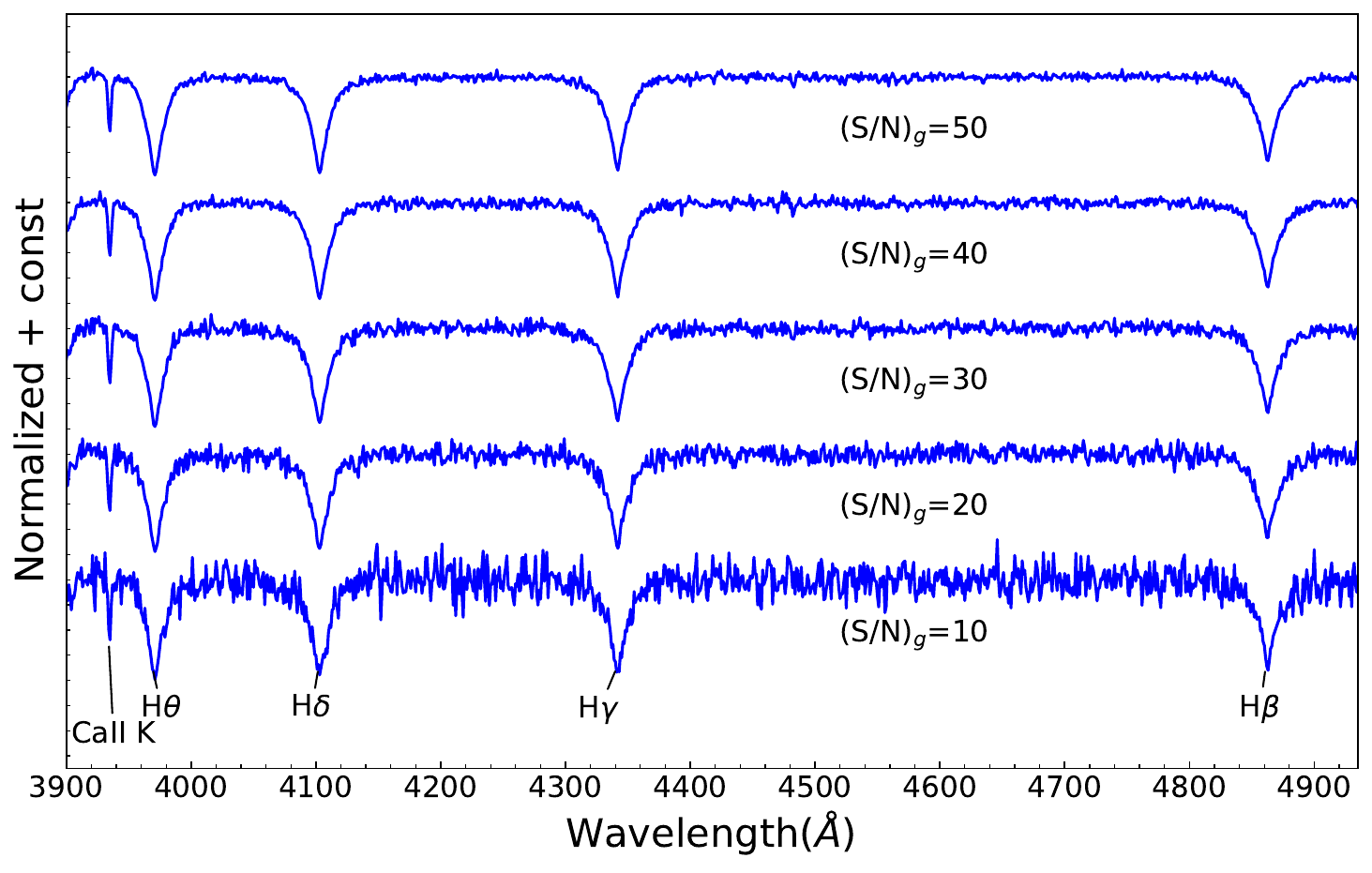}
    \plotone{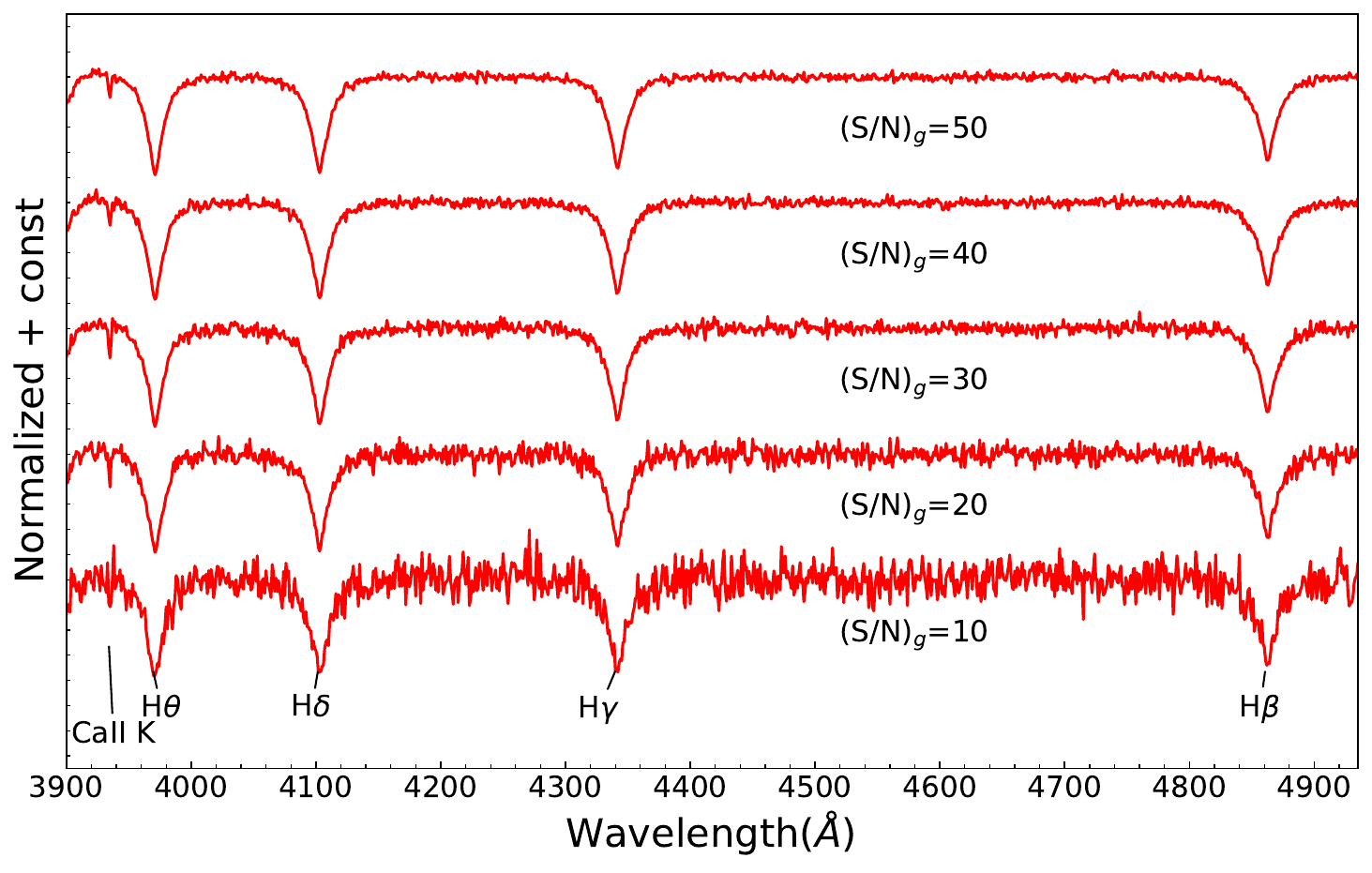}
    \plotone{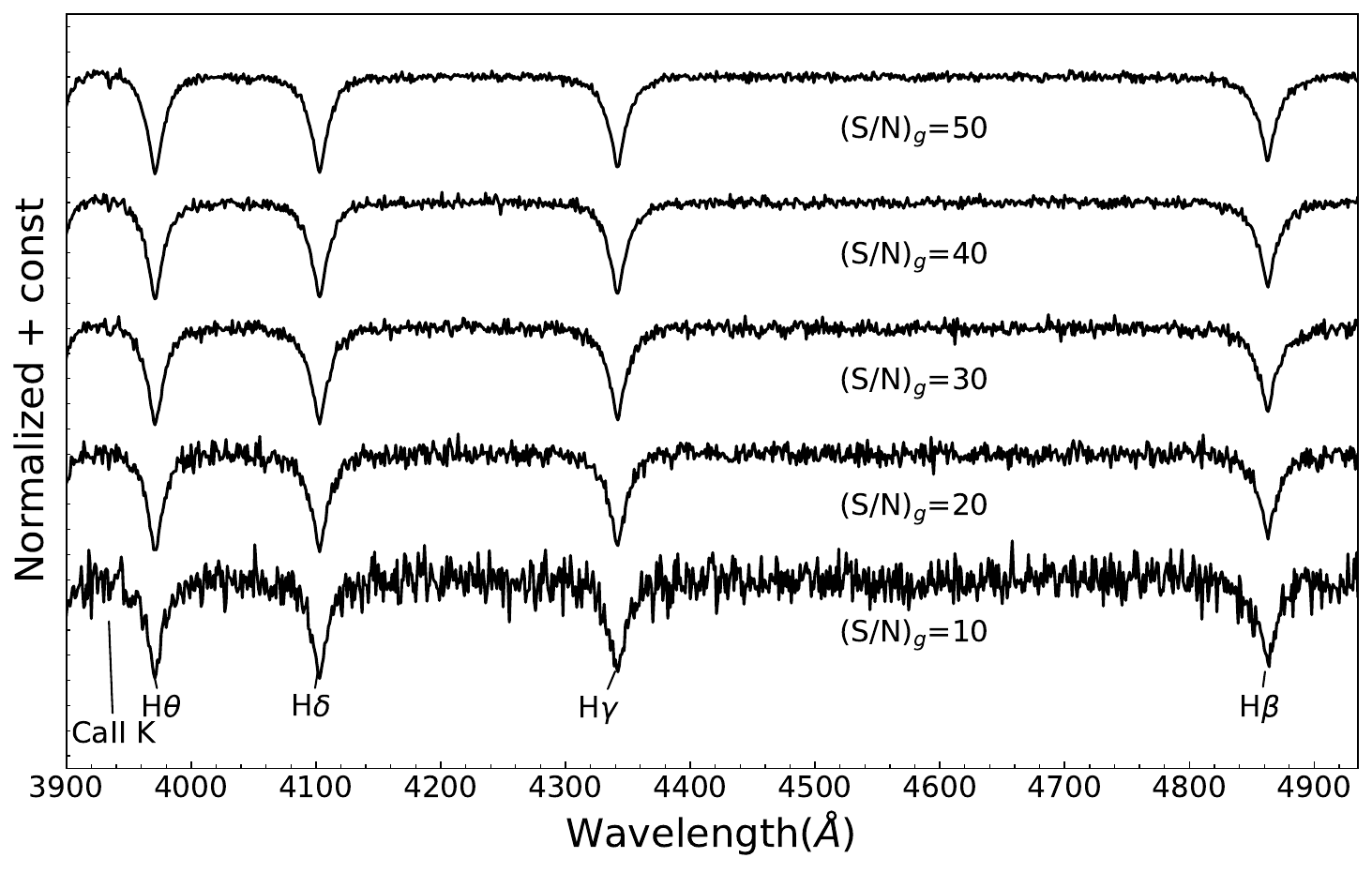}
    \caption{Synthetic spectra with $T_{\rm eff}= 9,000\,K$ and log\,$g=3.5$ at different  (S/N)$_g$ (10-50) and [Fe/H]. From top to bottom, blue represents [Fe/H]= -1, red represents [Fe/H]= -2, black represents [Fe/H]= -3.\label{fig:snr-l2}}
\end{figure}

\begin{figure*}
	\epsscale{1.2}
    \plotone{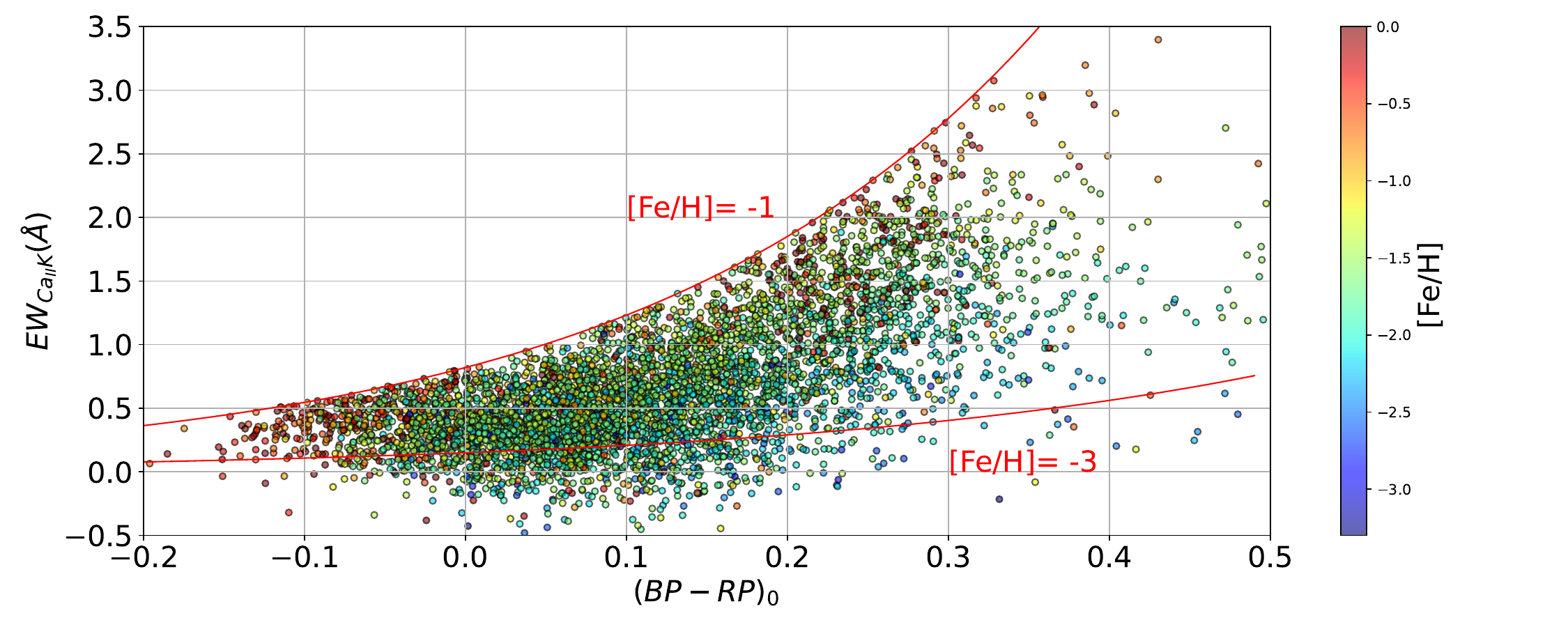}
    \caption{The distribution of our BHB sample in $(BP-RP)_0$ vs. $\mathrm{EW_{Ca\,II\,K}}$ plane. The colorbar indicates the metalicities of [Fe/H] of the BHB candidates. The metallicities are adopted from \citet{Xiang2022}. The red solid lines are the same in Figure~\ref{fig:Cak-feh-1}.\label{fig:5377-cak}}
\end{figure*}
During the process of selecting the BHB stars, the primary absorption lines utilized are \ion{Ca}{2}\,K line, \hdelta, \hgamma\, and $\rm{H}\beta$, which are notably strong in A-type stars. However, the noise and absorption line may degenerate together in cases where the  (S/N)$_g$ is low. To illustrate this, the synthetic spectra calculated by \citet{Allende2018} with
log\,$g=3.5$ and [Fe/H]\,$=-1.7$ from globular cluster BHB stars\citep{Wilhelm1999a} under various (S/N)$_g$ from 10 to 50 with the step of 10 are shown in Figure~\ref{fig:snr-ll}. The S/N is added using a Gaussian function. For the panels from top to bottom, the spectra represent $T_{\rm eff} = 8,000\,K, 9,000\,K$, and $10,000\,K$, respectively. It can be seen that the \ion{Ca}{2}\,K  line gradually weakens with the increase in temperature. At $T_{\rm eff} = 8,000\,K$, the $\mathrm{Ca\,II\,K}$ line is still relatively strong, and the  (S/N)$_g$ has minimal impact. However, as the temperature continues increasing, the \ion{Ca}{2}\,K line begins to weaken. At $T_{\rm eff} = 10,000\,K$ and  (S/N)$_g = 10$, noise has a more significant impact on the \ion{Ca}{2}\,K line, which could lead to superimposing of noise and the \ion{Ca}{2}\,K line, 
resulting uncertainties to the value of the calculated equivalent width.


Figure~\ref{fig:snr-l2} shows the synthetic spectra \citep{Allende2018} with $T_{\rm eff} = 9,000$\,K, log\,$g=3.5$ and [Fe/H] = -1, -2, and -3 (from top to bottom) under various (S/N)$_g$. The \ion{Ca}{2}\,K line is an important indicator of metallicities for BHB stars as under such temperatures the other meal lines are generally very weak or absent. With the reduction of (S/N)$_g$, noise and \ion{Ca}{2}\,K line are also superimposed, leading to large uncertainties of $\mathrm{EW_{Ca\,II\,K}}$ calculation in the spectra with (S/N)$_g=10$, especially for BHB stars with low metallicity, which usually have very weak $\mathrm{EW_{Ca\,II\,K}}$ lines. 
Based on Figure~\ref{fig:Cak-feh-1} and combining Figure~\ref{fig:snr-ll},\ref{fig:snr-l2}, we can see that the low 
(S/N)$_g$ may bring some uncertainties on the selection of BHB stars with high $T_{\rm eff}$ and low [Fe/H], but little or negligible impact on that for BHB stars with low temperatures or high [Fe/H], which usually have stronger  \ion{Ca}{2}\,K lines than above cases. However, for the effect of low signal-to-noise on the selection of BHB stars, a quantitative analysis is complicated due to the observational selection effect of LAMOST and the variation of the stellar formation rate (SFR) with different metallicities.

The stellar atmospheric parameters of \citet{Xiang2022} are determined by fitting the synthetic spectra with the observation one. Here, we use the $T_{\rm eff}$ and [Fe/H] estimated by \citet{Xiang2022} to simply estimate the purity of our BHB samples. From Figure~\ref{fig:snr-pure}, our purity is about 83.4\% even at (S/N)$_g=10$, which indicates that when using the \ion{Ca}{2}\,K line cuts the (S/N)$_g$ has a certain impact, but it is relatively small.

\subsection{Spatial distribution of our BHB stars}
\begin{figure}
    \epsscale{1.0}
	\plotone{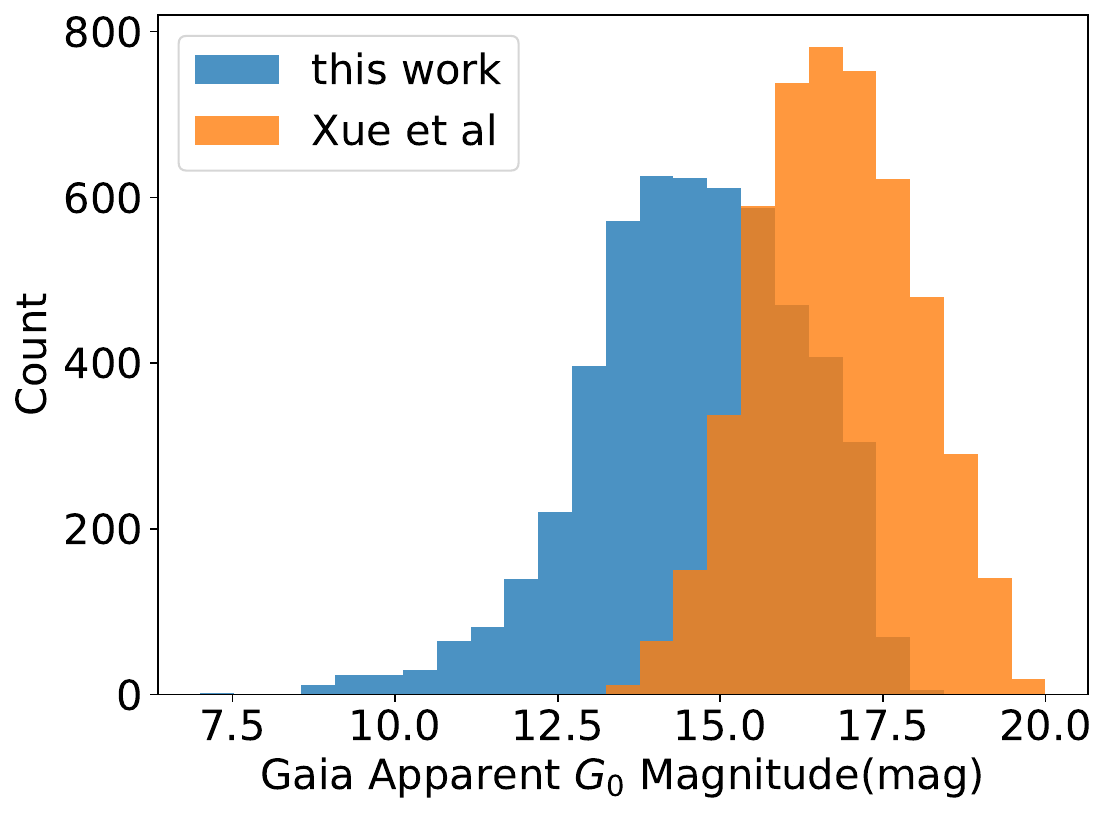}
	\caption{Distribution of apparent magnitudes for this work (blue) and \citet{Xue2011} (orange). \label{fig:g0}}
\end{figure}

BHB stars are deemed to have an almost constant absolute
magnitude. Thus, their apparent magnitudes can be used as a distance indicator, that is, more distant BHB stars correlated with larger apparent magnitudes. Figure~\ref{fig:g0} compares the Gaia DR3 $G$ magnitude of BHB stars from this work and \citet{Xue2011}, for which the extinction is corrected using \citet{Schlegel1998}. The peak value of $G$ magnitude of our sample and \citet{Xue2011} are about 14.6\,mag and 16.8\,mag, respectively. Compared to \citet{Xue2011} sample, our BHB stars are indeed brighter and thus should be closer. 

\begin{figure*}
    \epsscale{1.0}
	\plotone{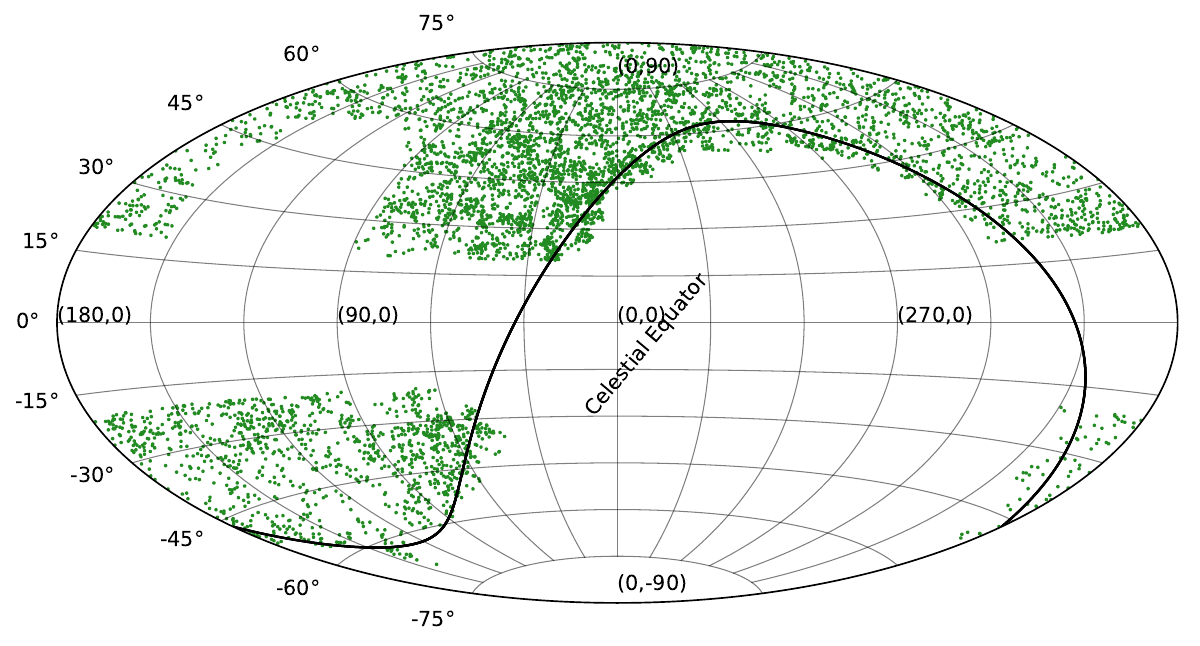}
	\plotone{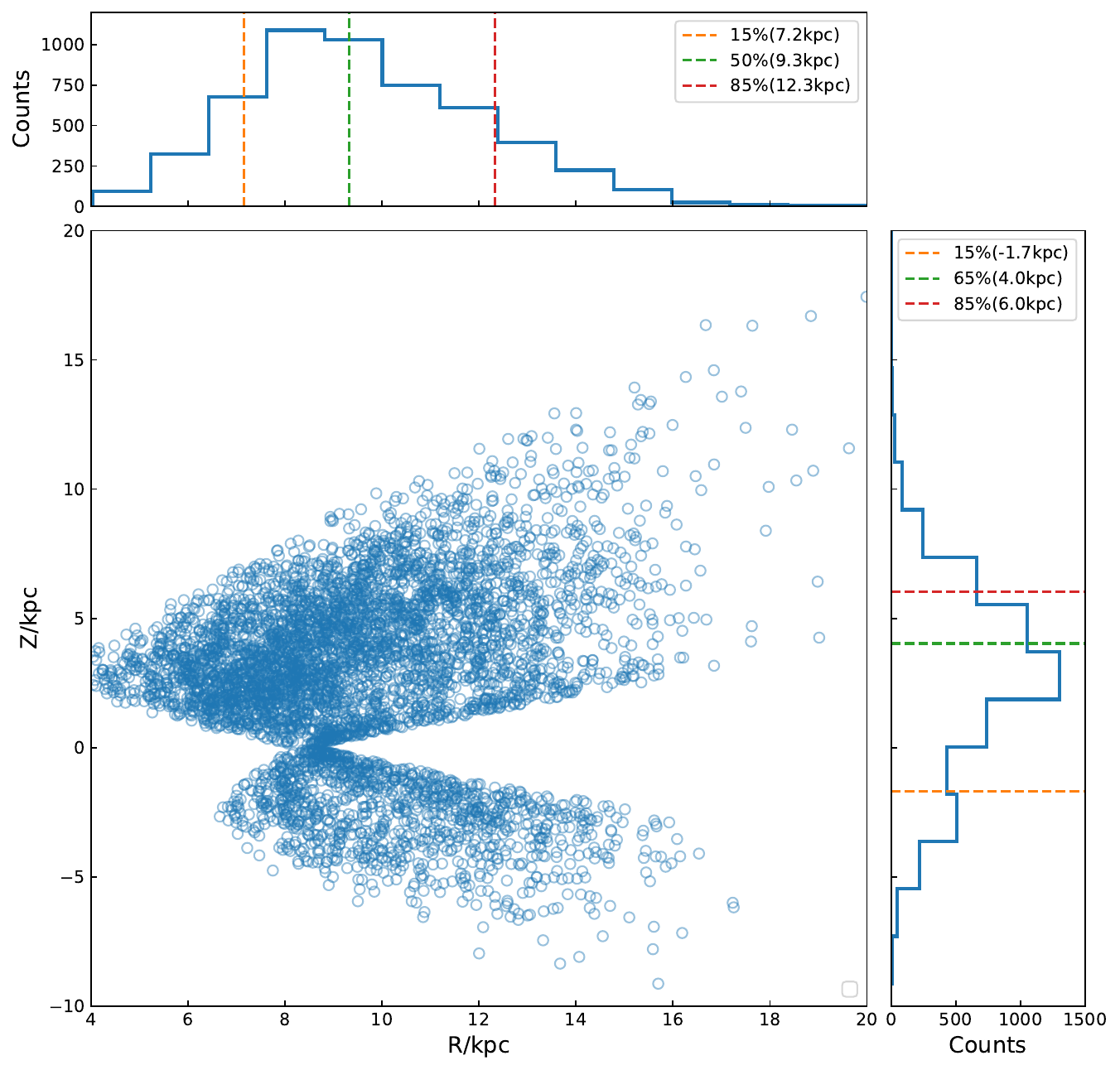}
	\caption{The top panel: the Galactic sky coverage of our BHB sample; the bottom panel: the spatial distribution in the $R-Z$ plane. In the top and right histograms, the orange, green, and red dashed lines represent 15{\%}, 50{\%}, and 85{\%} of R and Z. \label{fig:6848sp}}
\end{figure*}

Figure~\ref{fig:6848sp} displays the Galactic sky coverage and spatial distribution of our BHB stars.
It shows that most of our BHB stars are located within $\lvert Z \rvert<15$\,kpc. From the right histogram, we can see that the peak value of the $|Z|$ of our BHB stars is from 2 to 3\,kpc, where $\lvert Z \rvert<4$\,kpc is generally used to divide the locations of the halo and the thick disk for the Milky Way \citep{Norris1985, Beers2002}. In addition, there are about 35\% of our BHB stars with $|Z|>4$\,kpc, but about 85\% of them with $|Z|<6$\,kpc,  
 which indicates that most of our BHB stars are located in the inner halo or overlap with the disk location in the Milky Way. 
 It is worth pointing out that many BHB stars with a small $Z$ (the vertical distance to the Galactic middle disk) are also successfully selected out. Our sample not only complements the known BHB stars but also covers the volume closer to the Sun than the other known sample.
 Such a sample is essential for the BHB stars as a probe to study the properties of the Galactic halo, especially in the solar neighborhood. 


\section{Conclusions}

In this work, we successfully identify 5,355+81 BHB stars with $|Glat|>20^{\circ}$ and (S/N)$_g > 10$ from the LAMOST DR5. To get a clean sample of BHB stars, we combine the line indices cut, the \ddo\ method with the scale width-shape method, and the $\mathrm{EW}_\mathrm{Ca\,II\,K}$ cuts are also adopted to remove the metal-rich contaminations.

Based on the atmospheric parameters obtained by \citet{Xiang2022}, most of our BHB stars are in agreement with the range of BHB atmospheric parameters and the theoretical $T_{\rm eff}$-log\,$g$ evolutionary tracks of the BHB stars. It indicates the reliability of our BHB sample. But within 7,000-8,000\,K, there are about 350 BHB stars with large log\,$g$ around $4.5$. We also noted that these stars usually have overestimated metallicities, which will result in large uncertainties to the estimations of $T_\mathrm{eff}$ and log\,$g$. 
   
Our BHB samples are mainly concentrated in the inner halo or around the disk of the Milky Way. 
By applying the line indices of \ion{Ca}{2}\,K as the metallicity cuts, we successfully select out the rare halo BHB stars which are located in the Galactic disk region ($\lvert Z \rvert<4$\,kpc). Combined with \citet{Xue2011} samples, BHB stars can be used as a better probe to study the kinematics and structural characteristics of the Milky Way. 
\begin{acknowledgments}
This study is supported by the National Key Basic R$\&$D Program of China No. 2019YFA0405500; the National Natural Science Foundation of China under grant No. 12173013; the project of Hebei provincial department of science and technology under the grant number 226Z7604G, and the Hebei NSF (No. A2021205006, A2022205018). We also acknowledge the science research grants from the China Manned Space Project. The Guoshoujing Telescope (the Large Sky Area Multi-Object Fiber Spectroscopic Telescope LAMOST) is a National Major Scientific Project built by the Chinese Academy of Sciences. LAMOST is operated and managed by the National Astronomical Observatories, Chinese Academy of Sciences. 
\end{acknowledgments}

%

\vspace{5mm}
\facilities{LAMOST}


\software{astropy \citep{2013A&A...558A..33A},  
          }



\bibliographystyle{aasjournal}

\end{document}